\documentclass[acmsmall, nonacm, screen]{acmart}
\newif\ifcomments\commentsfalse
\usepackage[scaled=1.0]{DejaVuSansMono}
\usepackage[capitalise]{cleveref}
\usepackage{xspace}
\usepackage{amsfonts}
\usepackage{booktabs}
\usepackage{tcolorbox}
\usepackage{subcaption}
\usepackage{xcolor} 
\usepackage{mathtools}
\usepackage{stmaryrd}
\usepackage{listings}
\usepackage{framed}
\usepackage[edges]{forest}
\usepackage{tikz}
\usepackage{algorithm}
\usepackage{mathpartir}
\usepackage[noend]{algpseudocode}
\usepackage{wrapfig}
\usepackage{cprotect}

\crefname{section}{Sec.}{Secs.}
\usepackage{enumitem}
\usepackage{suffix}
\usepackage{langrules}

\newcommand{\DefineTypeRule}[3]{\DefineRule[#1Rule]{#1}{#2}{#3}}
\newcommand{\DefineEvalRule}[3]{\DefineRule[E#1Rule]{E-#1}{#2}{#3}}

\DefineTypeRule{Var}{\Gamma(x) = \tau}{\Gamma \proves x : \tau}

\DefineTypeRule{Arch}{ }{\Gamma \proves \Arch : \ArchT}

\DefineTypeRule{Gate}{ }{\Gamma \proves \Gate : \GateT}

\DefineTypeRule{State}{ }{\Gamma \proves \StateIn : \StateT}

\DefineTypeRule{Trans}{ }{\Gamma \proves \Trans : \TransT}
\DefineTypeRule{IdTrans}{ }{\Gamma \proves \programfont{IdTrans} : \TransT}

\DefineTypeRule{Map}{ }{\Gamma \proves \QubitMap : \Qubit \to \LocT}

\DefineTypeRule{Float}{ }{\Gamma \proves r : \Float}

\DefineTypeRule{Int}{ }{\Gamma \proves n : \Int}

\DefineTypeRule{String}{ }{\Gamma \proves \mathit{str} : \StringT}

\DefineTypeRule{True}{ }{\Gamma \proves \True : \Bool}

\DefineTypeRule{False}{ }{\Gamma \proves \False : \Bool}

\DefineTypeRule{Loc}{
  \Gamma \proves e : \Int
}{\Gamma \proves \loc(e) : \LocT}

\DefineTypeRule{Abs}{
  \Gamma,x\ty\tau_1 \proves e : \tau_2
}{\Gamma \proves \fun{x}{e} : \tau_1 \to \tau_2}

\DefineTypeRule{App}{
  \Gamma \proves e_1 : \tau_1 \to \tau_2 \\
  \Gamma \proves e_2 : \tau_1
}{\Gamma \proves \app{e_1}{e_2} : \tau_2}

\DefineTypeRule{List}{
  \Gamma \proves \overline{e} : \tau
}{\Gamma \proves [\overline{e}] : \List[\tau]}

\DefineTypeRule{ListAccess}{
  \Gamma \proves e_1 : \List[\tau] \\
  \Gamma \proves e_2 : \Int
}{\Gamma \proves e_1[e_2] : \tau}

\DefineTypeRule{If}{
  \Gamma \proves e_1 : \Bool \\
  \Gamma \proves e_2 : \tau \\
  \Gamma \proves e_3 : \tau
}{\Gamma \proves \ITE{e_1}{e_2}{e_3} : \tau}

\DefineTypeRule{ArithFloat}{
  \Gamma \proves e_1 : \Float \\
  \Gamma \proves e_2 : \Float
}{\Gamma \proves e_1 \otimes e_2 : \Float}

\DefineTypeRule{ArithInt}{
  \Gamma \proves e_1 : \Int \\
  \Gamma \proves e_2 : \Int
}{\Gamma \proves e_1 \otimes e_2 : \Int}

\DefineTypeRule{StructAccess}{
  \Gamma \proves e : S \\
  \mathrm{types}(S) = \overline{x}\ty\overline{\tau}
}{\Gamma \proves e.x_i : \tau_i}

\DefineTypeRule{Struct}{
  \mathrm{types}(S) = \overline{x}\ty\overline{\tau} \\
  \Gamma \proves \overline{e} : \overline{\tau}
}{\Gamma \proves S\{\overline{x} = \overline{e}\} : S}

\DefineTypeRule{Fun}{
  \mathrm{funtype}(F) = \overline{\tau_a} \to \tau \\
  \Gamma \proves \overline{e} : \overline{\tau_a}
}{\Gamma \proves F(\overline{e}) : \tau}

\DefineTypeRule{Pair}{
  \Gamma \proves e_1 : \tau_1 \\
  \Gamma \proves e_2 : \tau_2
}{\Gamma \proves (e_1, e_2) : \tau_1 \times \tau_2}

\DefineTypeRule{Proj}{
  \Gamma \proves e : \tau_1 \times \tau_2
}{\Gamma \proves \proj{i}{e} : \tau_i}

\DefineEvalRule{Ctx}{
  e \stepsto e'
}{E[e] \stepsto E[e']}

\DefineEvalRule{StructAccess}{ }{S\{\overline{x} = \overline{v}\}.x_i \stepsto v_i}

\DefineEvalRule{Arith}{
  v_1 \otimes v_2 = w
}{v_1 \otimes v_2 \stepsto w}

\DefineEvalRule{IfT}{ }{\ITE{\True}{e_1}{e_2} \stepsto e_1}

\DefineEvalRule{IfF}{ }{\ITE{\False}{e_1}{e_2} \stepsto e_2}

\DefineEvalRule{App}{ }{\app{(\fun{x}{e})}{v} \stepsto \subst{e}{x}{v}}

\DefineEvalRule{Proj}{ }{\proj{i}{(v_1, v_2)} \stepsto v_i}

\DefineEvalRule{ListAccess}{1 \leq i \leq n}{[v_1, \dotsc, v_{n}][i] \stepsto v_i}

\DefineEvalRule{Push}{ }{\Push([\overline{v}], w) \stepsto [\overline{v}, w]}

\DefineEvalRule{Concat}{ }{\Concat([\overline{v}], [\overline{w}]) \stepsto [\overline{v}, \overline{w}]}

\DefineEvalRule{ContainsT}{
  v = v_i
}{\programfont{contains}([\overline{v}], v) \stepsto \True}

\DefineEvalRule{ContainsF}{
  \forall i \in [0,n]\ldotp v_i \neq v
}{\programfont{contains}([v_1, \dotsc, v_n], v) \stepsto \False}

\DefineEvalRule{Map}{ }{\Map((\fun{x}{e}), [v_1, \dotsc, v_n]) \stepsto [\subst{e}{x}{v_0}, \dotsc, \subst{e}{x}{v_n}]}

\DefineEvalRule{FoldEmp}{ }{\Fold((\fun{x,y}{e}), v, []) \stepsto v}

\DefineEvalRule{FoldV}{n \geq 0}{\Fold((\fun{x,y}{e}), w, [v_0, \dotsc, v_n]) \stepsto \Fold((\fun{x,y}{e}), \subst*{e}{{x}{v_0}{y}{w}}, [v_1, \dotsc, v_n])}

\DefineEvalRule{Combinations}{1\leq i \leq n}{\programfont{combinations}([v_1, \dotsc, v_n], i) \stepsto [\overline{l}] \text{, $l_k$ is a list of $i$ elements from $\overline{v}$}}

\DefineEvalRule{Edges}{ }{\programfont{edges}((V,E)) \stepsto E}

\DefineEvalRule{EdgesBtwnT}{(u,v) \in E}{\programfont{edges\_between}((V,E), u, v) \stepsto [(u,v)] }

\DefineEvalRule{EdgesBtwnF}{(u,v) \not\in E}{\programfont{edges\_between}((V,E), u, v) \stepsto [] }

\DefineEvalRule{AllPaths}{ }{\programfont{all\_paths}(A, \overline{s}, \overline{t}, \overline{b}) \stepsto [\overline{p}] \text{, $p_k$ is a path from $s_i$ to  $t_j$ using no vertices in $b$}}

\DefineEvalRule{SteinerTree}{ }{\programfont{steiner\_trees}(A, \overline{s},  \overline{b}) \stepsto [\overline{t}] \text{, $t_k$ is a Steiner tree for $\overline{s}$ assuming vertices in $b$ are blocked}}

\DefineEvalRule{Qubits}{ }{\programfont{qubits}(\gate(\overline{q})) \stepsto \overline{q}}

\DefineEvalRule{GateType}{ }{\programfont{gate\_type}(\gate(\overline{q})) \stepsto \gate}

\DefineEvalRule{HorizE}{v_1 = 0, v_2 = 1}{\programfont{horizontal\_neighbors}(\loc(v_1), v_2)  \stepsto []}
\DefineEvalRule{HorizR}{v_1 \mod v_2 = 0, v_2 \neq 1}{\programfont{horizontal\_neighbors}(\loc(v_1), v_2)  \stepsto [\loc(v_1+1)]}
\DefineEvalRule{HorizL}{v_1 \mod v_2 = v_2 - 1, v_2 \neq 1}{\programfont{horizontal\_neighbors}(\loc(v_1), v_2)  \stepsto [\loc(v_1-1)]}
\DefineEvalRule{HorizB}{0 < v_1 \mod v_2 < v_2 - 1}{\programfont{horizontal\_neighbors}(\loc(v_1), v_2)  \stepsto [\loc(v_1-1), \loc(v_1+1)]}

\DefineEvalRule{VertE}{v_1 = 0, v_3 = 1}{\programfont{vertical\_neighbors}(\loc(v_1), v_2, v_3)  \stepsto []}
\DefineEvalRule{VertBe}{v_1 / v_2 = 0, v_3 \neq 1}{\programfont{vertical\_neighbors}(\loc(v_1), v_2, v_3)  \stepsto [\loc(v_1+v_2)]}
\DefineEvalRule{VertAb}{v_1 / v_2 = v_3-1, v_3 \neq 1}{\programfont{vertical\_neighbors}(\loc(v_1), v_2, v_3)  \stepsto [\loc(v_1-v_2)]}
\DefineEvalRule{VertB}{0 < v_1 / v_2 < v_3 - 1}{\programfont{vertical\_neighbors}(\loc(v_1), v_2, v_3)  \stepsto [\loc(v_1+v_2), \loc(v_1 - v_2)]}

\DefineEvalRule{TwoD}{ }{\programfont{to\_2d}(\loc(v_1), v_2) \stepsto (v_1 \mod v2, v_1 / v_2)}

\DefineEvalRule{ValSwapL}{ m(q) = \loc(v_1)}{\programfont{value\_swap}(m, \loc(v_1), \loc(v_2))(q) \stepsto \loc(v_2)}
\DefineEvalRule{ValSwapR}{ m(q) = \loc(v_2)}{\programfont{value\_swap}(m, \loc(v_1), \loc(v_2))(q) \stepsto \loc(v_1)}
\DefineEvalRule{ValSwapN}{ m(q) \not\in \{\loc(v_1), \loc(v_2)\}}{\programfont{value\_swap}(m, \loc(v_1), \loc(v_2))(q) \stepsto m(q) }

\DeclareMathOperator*{\argmax}{argmax}

\DeclareMathOperator*{\cx}{\textsf{cx}}
\usetikzlibrary{arrows.meta, positioning, hobby}

\newcommand{\proves}{\vdash}
\newcommand{\ty}{\mkern2mu\mathord{:}\mkern2mu}
\newcommand{\alt}{\mkern3mu\mid\mkern3mu}
\newcommand{\NN}{\mathbb{N}}
\newcommand{\RR}{\mathbb{R}}
\newcommand{\stepsto}{\longrightarrow}
\newcommand{\denote}[1]{\mathchoice%
  {\left\llbracket {#1} \right\rrbracket}
  {\llbracket {#1} \rrbracket}
  {\llbracket {#1} \rrbracket}
  {\llbracket {#1} \rrbracket}%
}

\makeatletter
\def\subst@inner#1#2#3\_nil{#1 \mapsto #2\ifx\relax#3\relax\else,\mkern2mu\subst@inner#3\_nil\fi}%
\newcommand{\raw@subst}[3]{
  \mathchoice{#1{\left[#2\subst@inner#3\_nil\right]}}
             {#1[#2\subst@inner#3\_nil]}
             {#1[#2\subst@inner#3\_nil]}
             {#1[#2\subst@inner#3\_nil]}%
}
\newcommand{\subst}[3]{\raw@subst{#1}{}{{#2}{#3}}}
\WithSuffix\newcommand\subst*[2]{\raw@subst{#1}{}{#2}}
\makeatother

\newcommand{\exec}{\ensuremath{E}\xspace}

\newcommand{\gates}{\textit{Gates}\xspace}
\newcommand{\indexedgates}{\ensuremath{\mathit{Instrs}}\xspace}
\newcommand{\gate}{\ensuremath{g}\xspace}

\newcommand{\transition}{\ensuremath{t}\xspace}
\newcommand{\transrel}{\ensuremath{\rightarrow}\xspace}
\newcommand{\circuit}{\ensuremath{C}\xspace}

\newcommand{\locs}{\textit{Locs}\xspace}
\newcommand{\devinstruct}{\textit{DInstrs}\xspace}
\newcommand{\qubits}{\textit{Qubits}\xspace}

\newcommand{\map}{\textsf{map}\xspace}
\newcommand{\amap}{\map}
\newcommand{\cost}{\ensuremath{c}\xspace}
\newcommand{\realizable}{\ensuremath{\mathit{Real}}\xspace}
\newcommand{\true}{\textsf{true}\xspace}

\newcommand{\route}{\textsf{routes}}
\newcommand{\ins}{\mathit{ins}}
\newcommand{\Vars}{\ensuremath{\mathcal{V}}\xspace}

\newcommand{\routedgates}{\programfont{routed\_gates}}
\newcommand{\costFun}{\programfont{cost}}

\newcommand{\swap}{\textsc{swap}\xspace}
\newcommand{\cnot}{\textsc{cx}\xspace}
\newcommand{\T}{\ensuremath{T}\xspace}
\newcommand{\qmr}{\textsc{qmr}\xspace}
\newcommand{\scmr}{\textsc{scmr}\xspace}
\newcommand{\nisq}{\textsc{nisq}\xspace}
\newcommand{\nisqmr}{\textsc{nisqmr}\xspace}
\newcommand{\nisqve}{\textsc{nisq-ve}\xspace}
\newcommand{\raa}{\textsc{raa}\xspace}
\newcommand{\mqlss}{\textsc{mqlss}\xspace}
\newcommand{\ilq}{\textsc{ilq}\xspace}
\newcommand{\tiqmr}{\textsc{tiqmr}\xspace}
\newcommand{\sat}{\textsc{sat}\xspace}
\newcommand{\solver}{Amaro\xspace}

\newcommand{\lang}{Amaro\xspace}

\newcommand{\framework}{Amaro\xspace}
\newcommand{\qiskit}{\textsc{{qiskit}}\xspace}

\newcommand{\qsynth}{\textsc{{qsynth2}}\xspace}
\newcommand{\dascot}{\textsc{dascot}\xspace}
\newcommand{\sabre}{\textsc{sabre}\xspace}
\newcommand{\enola}{\textsc{enola}\xspace}
\newcommand{\shaper}{\textsc{shaper}\xspace}

\newcommand{\qpu}{\textsc{qpu}\xspace}
\newcommand{\qec}{\textsc{qec}\xspace}
\newcommand{\qpus}{\textsc{qpu}{\footnotesize s}\xspace}
\newcommand{\solution}{\mathit{Sol}\xspace}
\newcommand{\prog}{\mathit{P}\xspace}
\newcommand{\snew}{s_{\mathit{new}}}
\newcommand{\scurr}{s_{\mathit{curr}}}
\newcommand{\temp}{\tau}
\newcommand{\tinit}{\tau_i}
\newcommand{\tfinal}{\tau_f}

\newcommand{\dr}{DR}
\newcommand{\graph}{A}
\newcommand{\interactgraph}{\ensuremath{\mathcal{I}}}
\definecolor{codegreen}{rgb}{0,0.6,0}
\definecolor{codegray}{rgb}{0.5,0.5,0.5}
\definecolor{codepurple}{rgb}{0.58,0,0.82}
\definecolor{backcolour}{rgb}{0.95,0.95,0.92}
\definecolor{myYellow}{HTML}{E69F00} 
\definecolor{myGreen}{HTML}{009E73}
\definecolor{myPurple}{HTML}{CC78A7}
\definecolor{myOrange}{HTML}{D55E00}
\definecolor{mygray}{rgb}{0.98,0.98,0.98}
\definecolor{commentcolor}{RGB}{0,100,0}
\definecolor{shadecolor}{named}{mygray}
\lstdefinelanguage{QRML}{
  keywords={GateRealization, Transition},
  keywordstyle=\color{blue!60!green!80!black},
  keywords=[2]{State, Arch, Gate, Trans, QubitMap},
  keywordstyle=[2]\color{red!60!blue!80!black},
  keywords=[3]{Loc, Int, Float, List, Bool},
  keywordstyle=[3]\color{myOrange},
  identifierstyle=\color{black},
  sensitive=false,
  comment=[l]{//},
  morecomment=[s]{/*}{*/},
  commentstyle=\color{commentcolor}\ttfamily,
  stringstyle=\color{red}\ttfamily,
  morestring=[b]',
  morestring=[b]"
}

\newenvironment{mybox}[1][gray!10]{  
    \begin{tcolorbox}[ 
        left=0pt,
        right=0pt,
        top=0pt,
        bottom=0pt,
        colback=#1,
        colframe=#1,
        width=0.99\dimexpr\columnwidth\relax,
        boxsep=2pt,
        arc=0pt,outer arc=0pt,
    ]
}{
    \end{tcolorbox}
}

\newcommand{\programfont}[1]{\texttt{\scriptsize{#1}}\xspace}
\newcommand{\StructFont}[1]{\programfont{\color{blue!60!green!80!black}#1}}
\newcommand{\ConstFont}[1]{\programfont{\color{red!60!blue!80!black}#1}}
\newcommand{\Gate}{\ConstFont{Instr}}
\newcommand{\Arch}{\ConstFont{Arch}}
\newcommand{\Trans}{\ConstFont{Trans}}
\newcommand{\StateIn}{\ConstFont{State}}
\newcommand{\GateRealization}{\StructFont{GateRealization}}
\newcommand{\QubitMap}{\ConstFont{QubitMap}}
\newcommand{\Transition}{\StructFont{Transition}}

\newcommand{\True}{\programfont{true}}
\newcommand{\False}{\programfont{false}}
\newcommand{\IfN}{\programfont{if}}
\newcommand{\ThenN}{\programfont{then}}
\newcommand{\ElseN}{\programfont{else}}
\newcommand{\ITE}[3]{\IfN~{#1}~\ThenN~{#2}~\ElseN~{#3}}
\newcommand{\fun}[2]{|{#1}| \to {#2}}
\newcommand{\app}[2]{{#1}~{#2}}
\newcommand{\loc}{\programfont{loc}}

\newcommand{\proj}[2]{\operatorname{\programfont{proj}}_{#1} {#2}}

\newcommand{\realizegate}{\programfont{realize\_gate}}
\newcommand{\availtrans}{\programfont{get\_transitions}}
\newcommand{\getlocations}{\programfont{get\_locations}}
\newcommand{\apply}{\programfont{apply}}

\newcommand{\Push}{\programfont{push}}
\newcommand{\Map}{\programfont{map}}
\newcommand{\Fold}{\programfont{fold}}
\newcommand{\Concat}{\programfont{concat}}

\newcommand{\TpFont}[1]{\programfont{\color{myOrange}#1}}
\newcommand{\LocT}{\TpFont{Loc}}
\newcommand{\Int}{\TpFont{Int}}
\newcommand{\Float}{\TpFont{Float}}
\newcommand{\Bool}{\TpFont{Bool}}
\newcommand{\List}{\TpFont{List}}

\newcommand{\ArchT}{\TpFont{ArchT}}
\newcommand{\StateT}{\TpFont{StateT}}
\newcommand{\StructT}{\mathit{S}}
\newcommand{\TransT}{\Transition}
\newcommand{\GateT}{\TpFont{InstrT}}
\newcommand{\Qubit}{\TpFont{Qubit}}
\newcommand{\GateRealT}{\GateRealization}
\newcommand{\StringT}{\TpFont{String}}

\setcopyright{cc}
\setcctype{by}
\acmDOI{10.1145/3776720}
\acmYear{2026}
\acmJournal{PACMPL}
\acmVolume{10}
\acmNumber{POPL}
\acmArticle{78}
\acmMonth{1}

\begin{document}
\title{Generating Compilers for Qubit Mapping and Routing}
\author{Abtin Molavi}
\orcid{0009-0006-1841-9565}
\affiliation{%
  \institution{University of Wisconsin-Madison}
  \country{USA}
}
\email{amolavi@wisc.edu}

\author{Amanda Xu}
\orcid{0009-0008-2279-5816}
\affiliation{%
  \institution{University of Wisconsin-Madison}
  \country{USA}
}
\email{axu44@wisc.edu}
\author{Ethan Cecchetti}
\orcid{0000-0001-7900-8328}
\affiliation{%
  \institution{University of Wisconsin-Madison}
  \country{USA}
}
\email{cecchetti@wisc.edu}

\author{Swamit Tannu}
\orcid{0000-0003-4479-7413}
\affiliation{%
  \institution{University of Wisconsin-Madison}
  \country{USA}
}
\email{stannu@wisc.edu}

\author{Aws Albarghouthi}
\orcid{0000-0003-4577-175X}
\affiliation{%
  \institution{University of Wisconsin-Madison}
  \country{USA}
}
\email{aws@cs.wisc.edu}

\begin{abstract}

To evaluate a quantum circuit on a quantum processor, one must find a mapping from circuit qubits to processor qubits and plan the instruction execution while satisfying the processor's constraints.
This is known as the \emph{qubit mapping and routing} (\qmr) problem.
High-quality \qmr solutions are key to maximizing the utility of scarce quantum resources and minimizing the probability of logical errors affecting computation.
The challenge is that the landscape of quantum processors is incredibly diverse and fast-evolving.
Given this diversity, dozens of papers have addressed the \qmr problem for different qubit hardware, connectivity constraints, and quantum error correction schemes by a developing a new algorithm for a particular context.
We present an alternative approach: automatically generating qubit mapping and routing compilers for arbitrary quantum processors.
Though each \qmr problem is different, we identify a common core structure---\emph{device state machine}---that we use to formulate an \emph{abstract} \qmr \emph{problem}.
Our formulation naturally leads to a compact domain-specific language for specifying \qmr problems and a powerful parametric algorithm that can be instantiated for any \qmr specification.
Our thorough evaluation on case studies of important \qmr problems shows that generated compilers are competitive with handwritten, specialized compilers in terms of runtime and solution quality. 

\end{abstract}

\maketitle 

\section{Introduction}
Quantum computation promises to surpass classical methods in important domains, potentially
unlocking breakthroughs in materials science, chemistry, machine learning, and beyond. %
Quantum computing is at an inflection point: scientists are scaling quantum hardware~\cite{IBM_Roadmap,Google_Roadmap}, %
demonstrating practical quantum error correction protocols~\cite{Google_Quantum_AI_and_Collaborators2024-fp}, and exploring promising application domains~\cite{kim2023evidence}. %
However, to fully realize the potential of quantum hardware available today and on the horizon, we need optimizing quantum circuit compilers. %
A compiler must convert architecture-independent, circuit-level descriptions of quantum programs to a form executable on a target \emph{quantum processing unit} (\qpu). %
Inefficient compilation that induces significant runtime overhead is unacceptable. %
For one, access to quantum compute is limited and costly. Further, quantum computation is error-prone, and longer computations are associated with a higher probability of a logical error, even when quantum error-correcting codes are applied. %

To enable execution of a quantum circuit on a target \qpu, a compiler must find a \emph{mapping} from circuit qubits to physical locations on the \qpu and plan the \emph{routing} of quantum instructions (gates) in a way that is compliant with the \qpu's physical and logical constraints. %
This is known as the \emph{qubit mapping and routing problem} (\qmr). 

The challenge for compiler designers is that the landscape of target quantum architectures is incredibly diverse and fast-evolving. %
First, there are several competing hardware realizations of an individual physical qubit, such as superconducting circuits, neutral atoms, trapped ions, and photons. %
Each physical realization of qubits imposes its own unique constraints on \qmr. 
For example, superconducting qubits are fixed in place, while neutral atom qubits can be shuttled in physical space. %
Second, qubits can be arranged and connected in a variety of ways. %
For example, superconducting circuits can be arranged in a linear array or a grid.
Third, going up the abstraction ladder, \emph{quantum-error correction} (\qec) schemes encode a logical qubit using several physical qubits, and each \qec scheme imposes its unique architectural constraints.
%



\begin{figure}[t]
    \centering
    \includegraphics[width=1.0\linewidth]{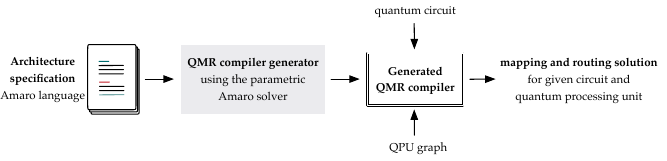}
    \vspace{-2em}
    \caption{Overview of our approach to \qmr compiler generation. Given a specification of a \qmr problem in the \lang language, we generate a compiler by instantiating the parametric \framework solver. The generated compiler takes a quantum circuit and target \qpu graph as input and produces a mapping and routing solution.}
    \label{fig:qmr-overview}
\end{figure}

The diversity in quantum architectures yields an array of \qmr problems to tackle for different qubit hardware, connectivity constraints, and quantum error correction schemes.
A recent survey \cite{QMR_Survey} explicitly enumerates dozens of papers which address a variant of the \qmr problem as their primary focus.
\cref{tab:qmr-problems} shows a selection of \qmr problems studied in the literature, highlighting the diversity of the considered constraints.
For each new set of mapping and routing constraints, 
researchers establish hardness results, identify connections to graph-theoretic problems, and develop specialized compilation algorithms. %
Ideally, we would prefer to avoid restarting this process from scratch for each new emerging architecture. %

Therefore, in this work we ask the following question:

\begin{center}
    \emph{Can we automatically synthesize a compiler from a specification of architectural constraints?}
\end{center}
To this end, we construct a framework called \framework (Abstract MApping and ROuting) that unifies and generalizes \qmr problems, illustrated in \cref{fig:qmr-overview}. %
Though each \qmr problem is different, we identify a common core structure that we use to define an \emph{abstract} \qmr \emph{problem}.
Then, each architecture-specific problem is a different concrete instantiation of the abstract problem. %
Our abstract \qmr problem is based on the view of a mapping and routing solution
as a sequence of \emph{device states}. 
A device state captures the current qubit mapping and which gates are evaluated in parallel at a given execution time step.
The architectural constraints dictate which
states and transitions between states are valid.

Our formulation of the abstract \qmr problem naturally leads to the \framework language, a domain-specific language for specifying \qmr problems.
A program in the \lang language  defines a \emph{device state machine} that describes the \qpu's physical and logical constraints.
With a few lines of code, we can specify a new \qmr problem and automatically generate a compiler for it.
For example, the definition of the most well-studied \qmr problem, for \emph{noisy intermediate-scale quantum} (\nisq) architectures, is just 12 lines in the \framework language (\cref{tab:qmr-problems} shows a selection of \qmr problems that we use as case studies along with line counts).

We demonstrate that our abstract \qmr problem can be solved with a simple parametric solver that can be instantiated for any \qmr problem.
Given a specification of a \qmr problem in the \framework language, we automatically generate a compiler for it by instantiating the parametric \framework solver.
Generally, \qmr problems have been shown to be \textsc{np}-hard \cite{qubitallocationsirachi,dascot} and constraint-based solutions have been shown to be unscalable. 
Our solver, instead, is approximate in nature.
It constructs the mapping and routing solution by incrementally building a sequence of device states, attempting to maximize the number of gates that can be executed in parallel.
We call this approach \emph{maximal-state construction} and study the properties of \qpus for which our solver can find maximal states.
We identify two desirable properties, \emph{monotonicity} and \emph{non-interference}.
It turns out, for all practical \qpus, we can find maximal states (via monotonicity), and for some \qpus (like in the \nisq setting), we can find \emph{maximum} states without requiring search (via non-interference)---speeding up the algorithm.
For managing the combinatorial explosion, we use the classic and simple \emph{simulated annealing} algorithm~\cite{kirkpatrick1983simulatedannealing}.
Our solver is suitable for industrial-scale quantum circuit compilation because it is simple, highly configurable, and amenable to 
parallelization.

We evaluate our approach with several case studies of important \qmr problems considered in prior work (see \cref{tab:qmr-problems}), including very recently introduced \qmr problems for fault-tolerant quantum computers~\cite{ilq,mqlss}.
Qualitatively, our results demonstrate the generality and versatility of our abstract \qmr formulation:
we are able to concisely specify \qmr problems for noisy and fault-tolerant quantum architectures on a variety of hardware realizations.
Quantitatively, we perform an experimental evaluation on a comprehensive circuit benchmark suite to assess the performance of our solver. %
Our results indicate that our generated compilers are competitive with handwritten compilers in terms of runtime and solution quality (details in \cref{sec:eval}).
For example, our solver finds solutions with the same cost or better than the 
leading industrial toolkit, \qiskit \cite{qiskit2024}, for half of our benchmarks. On some \qmr 
domains, we even outperform the prior state-of-the-art. For the case of interleaved logical qubits, 
our solutions are strictly higher-quality than the baseline for 93\% of cases.
We envision that our approach will simplify development of quantum compilers for the many new and emerging quantum architectures.
\begin{table}[t]
  \small
  \caption{A selection of \qmr problems that we use as case studies}
  \vspace{-0.5em}
  \label{tab:qmr-problems}
    \begin{tabular}{lll}
        \toprule
        Case study  & \lang LoC & Prior work \\
        \midrule
        Near-term superconducting (\nisqmr)  & 12 & \cite{sabre,mqtmap,cowtan2019qubit, shaik_et_al:LIPIcs.SAT.2024.26, Molavi2022QubitMA}  \\
        Near-term superconducting with variable error (\nisqve)    & 30 & \cite{notallqubits,noise-adaptive-murali} \\
        Trapped-ion compilation (\tiqmr)  & 51 & \cite{bach2025efficientcompilationshuttlingtrappedion, qccedsim}\\
        Reconfigurable atom array compilation (\raa) & 116 &\cite{enola, atomique, olsq-dpqa}\\
        Surface code mapping and routing (\scmr) & 40 &\cite{dascot, autobraid}\\ 
        Multi-qubit lattice surgery scheduling (\mqlss) & 56 &\cite{mqlss}\\
        Interleaved logical qubit compilation (\ilq) & 44 &\cite{ilq}\\
         \bottomrule
     \end{tabular}
\end{table}

To summarize, our contributions are the following:
\begin{itemize}
    \item An abstract formulation of the  mapping and routing problem, based on device state machines, that presents a uniform way of thinking of the zoo of \qmr problems (\cref{sec:prob-def}).
    \item A specification language for \qmr problems that enables concise expression of the unique constraints of a particular architecture family. (\cref{sec:lang})
    \item A powerful parametric solver that can be automatically instantiated into a compiler for a given \qmr problem from a specification (\cref{sec:solver})
    \item An extensive empirical evaluation demonstrating the generality of our approach and the quality of the synthesized compilers in comparison to handwritten compilers. (\cref{sec:case-studies,sec:eval})
\end{itemize}

\section{Background and Overview of our Approach}
In this section, we begin by giving an overview of quantum circuits, providing the relevant background for this work to readers unfamiliar with quantum computing.
We then introduce the \qmr problem through two key examples of target \qpus.
By studying these two examples, we highlight the commonalities and motivate the abstract \qmr problem and our specification language.

\subsection{Quantum Circuits}

We give a brief overview of quantum circuits.
Since we are interested in the mapping and routing problem, it suffices to consider the structure of quantum circuits and not their full semantics.
A comprehensive introduction can be found in any standard quantum computing text (e.g., \cite{Nielsen_Chuang_2010}).

\paragraph{Qubits} The unit of data in quantum computing is  called the \emph{qubit}. A qubit can be one of the two \emph{computational
basis} states, 0 and 1, or a linear combination of the two, with complex-number coefficients called \emph{amplitudes}.
The state of $n$ qubits is described by $2^n$ amplitudes.

\paragraph{Gates} Quantum states are transformed by operations called \emph{quantum gates}. In this work,
we focus on single-qubit and two-qubit gates. One important single-qubit gate is the \T gate. 
The \T gate leaves the 0 state unchanged, and applies a \emph{phase-shift} to the 1 state, multiplying its amplitude by $e^{i\pi/4}.$
The \T gate plays an important role in fault-tolerant quantum computation. 
While necessary for universality---the ability to approximate any quantum computation to arbitrary precision---\T gates are typically expensive to implement in the context of quantum error correction.

A common two-qubit gate is the \cnot gate. The \cnot gate is named for its action on the computational basis states 
as a ``conditional-not.'' If the first argument, called the \emph{control}, is 1, the \cnot applies a \textsc{not} 
to the second argument, which is called the \emph{target}. If the control is in the 0 state, the gate has no effect.
Another important two-qubit gate is the \textsc{swap} gate, which swaps the values of two qubits.

\begin{figure}[b]
  \centering
  \includegraphics[width=0.2\linewidth]{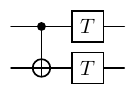}
  \caption{A simple quantum circuit}
  \label{fig:circuit-ex}
\end{figure}

\paragraph{Circuits} Quantum gates can be composed to produce a quantum circuit. \cref{fig:circuit-ex}
shows a simple quantum circuit. This example merely demonstrates the structure of circuits; the computation it performs is not significant. 
Each horizontal wire represents a qubit---two qubits are present in this example---and the circuit is read from left to right.
In this circuit, we first 
apply a \cnot gate to the two qubits (the top qubit is the control and the bottom qubit is the target), then a \T gate to each qubit. 
Equivalently, this circuit can be written as a sequence of instructions:
$\cnot~q_1~q_2; \T~q_1; \T~q_2$.

\subsection{An Introduction to \qmr Problems}
\label{sec:examples}
In this section, we introduce qubit mapping and routing through two key examples of target \qpus.
We chose two ostensibly very different \qmr problems, one targeting noisy quantum computers and the other targeting fault-tolerant quantum computers.
In comparing these two problems, we highlight a shared core structure with problem-specific parametric components.

Both \qmr problems reduce to finding an execution plan for a given circuit on a given \qpu.
We can express an execution plan as a sequence of  \emph{device states} of the \qpu. 
Intuitively, each state in the sequence represents a time step of execution. 
The differences between \qmr problems emerge as constraints that state sequences must satisfy to be considered valid solutions and \emph{cost} functions defining solution optimality.

We describe our two example \qmr problems with an eye towards this unifying perspective, summarizing
each problem in terms of the constraints imposed on the state sequences.

\subsubsection{Noisy-Intermediate Scale Quantum Computers} First, we consider mapping and routing for noisy-intermediate scale quantum (\nisq) architectures. 
These \qpus consist of up to hundreds of physical qubits and do not implement error-correction.
We will call this the \nisqmr problem. 
\nisqmr is well-studied with a variety of proposed solutions appearing in the literature, ranging from
greedy heuristic maximization \cite{sabre,cowtan2019qubit} to $A^*$ search \cite{zulehner2018efficient} to  reductions to satisfiability \cite{olsq2,satmapper,Molavi2022QubitMA}.

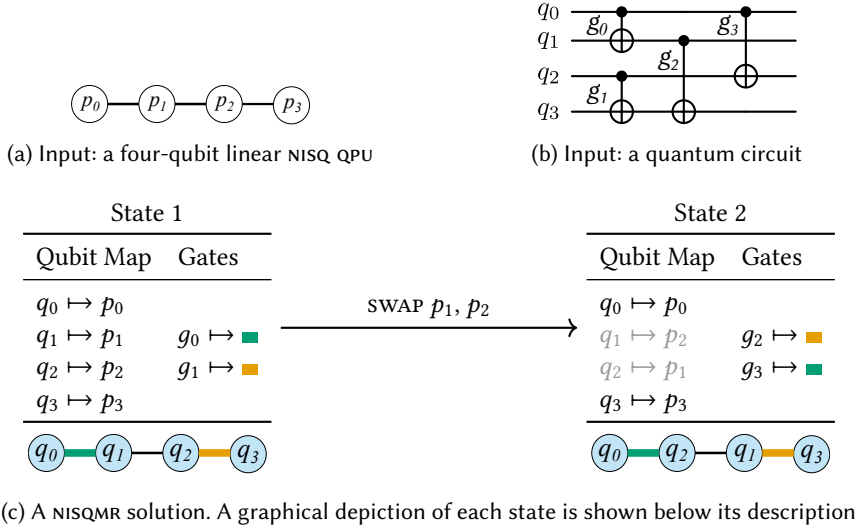
\begin{figure}
    \begin{subfigure}{0.45\linewidth}
        \centering
        \includegraphics{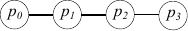}
        \caption{Input: a four-qubit linear \nisq \qpu }
        \label{fig:conn-graph}
    \end{subfigure}
    \begin{subfigure}{0.45\linewidth}
        \centering
        \includegraphics{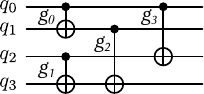}
        \caption{Input: a quantum circuit}
        \label{fig:circ}
    \end{subfigure}
\par\bigskip
    \begin{subfigure}{\linewidth}
      \centering
    \begin{tikzpicture}

      \node (step1-table) at (0,0) {
        \begin{tabular}{ll}
          \toprule
          Qubit Map & Gates \\
          \midrule
          $q_0 \mapsto p_0$ & \\
          $q_1 \mapsto p_1$ & $g_0 \mapsto \textcolor{myGreen}{\rule{6pt}{4pt}}$ \\
          $q_2 \mapsto p_2$ & $g_1 \mapsto\textcolor{myYellow}{\rule{6pt}{4pt}}$ \\
          $q_3 \mapsto p_3$ & \\
          \bottomrule
        \end{tabular}
      };

      \node[below=-0.1cm of step1-table] (step1-fig){
        \includegraphics{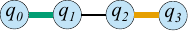}
      };
      \node[above=-0.1cm of step1-table] (step1-label){
        State 1
      };
  
      \node (step2-table) at (7.5,0) {
          \begin{tabular}{ll}
          \toprule
          Qubit Map & Gates \\
          \midrule
          $q_0 \mapsto p_0$ & \\
          \color{gray!80} $q_1 \mapsto p_2$ & $g_2 \mapsto \textcolor{myYellow}{\rule{6pt}{4pt}}$ \\
          \color{gray!80} $q_2 \mapsto p_1$ &$g_3 \mapsto \textcolor{myGreen}{\rule{6pt}{4pt}}$ \\
          $q_3 \mapsto p_3$ & \\
          \bottomrule
          \end{tabular}
      };

        \node[below=-0.1cm of step2-table] (step2-fig){
        \includegraphics{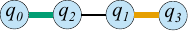}
      };

        \node[above=-0.1cm of step2-table] (step2-label){
        State 2
      };
  
      \draw[->, thick] (step1-table.east) -- node[above]{\swap $p_1$, $p_2$}(step2-table.west);
  
    \end{tikzpicture}
    \caption{A \nisqmr solution. A graphical depiction of each state is shown below its description}
    \label{fig:nisq-sol}
  \end{subfigure}
    \caption{Overview of the \nisqmr problem}
    \label{fig:nisq-conc}
\end{figure}

\paragraph{Mapping and Routing for \nisq Devices}
Because of spatial limitations, many \nisq \qpus, like the devices developed by IBM \cite{ibmqpus} and Google \cite{sycamorespecsheet,willowspecsheet}, only support two-qubit gates between certain pairs of qubits. 
We can represent a particular \qpu with a \emph{connectivity graph}. The connectivity graph includes an edge between
a pair of qubits if and only if the \qpu supports a two-qubit gate between them. A simple connectivity graph is shown in \cref{fig:conn-graph} (a linear graph where adjacent physical qubits are connected by an edge).
The goal of the compiler is to find a \emph{qubit map} from the qubits which appear in the circuit to the physical qubits of the \qpu such that two-qubit gates are executable, which is to say that the circuit qubits the gate acts on are mapped to adjacent qubits. For example, suppose we wish to execute the circuit in 
\cref{fig:circ} on this \qpu. The first gate, $g_0$, is between qubits $q_0$ and $q_1$, so we choose a map that maps these to a pair of adjacent physical qubits,
such as $p_0$ and $p_1$. Likewise, to execute the second gate, $g_1$, we map $q_2$ and $q_3$ to
$p_2$ and $p_3$. 
However, we cannot execute either of the remaining gates, $g_2$ and $g_3$, with this qubit map because there is no edge between the qubits the gates act on. 

\paragraph{Changing the Qubit Map via \swap Gates}
There is often no single static map such that all two-qubit gates are executable. 
Instead, the map is transformed over the course of execution by the insertion of \swap gates. 
\swap gates exchange the states of two adjacent qubits. In our example, to execute the \cnot gates $g_2$ and $g_3$ ($\cnot~q_1~q_3$ and $\cnot~q_0~q_2$),  we insert a \swap operation $\swap~p_1 ~p_2$.

A representation of our full mapping and routing solution is depicted in \cref{fig:nisq-sol}. 
Each box represents the \emph{device state} of the \qpu at a particular time step; each state is an assignment of some subset of the gates of the circuit to an edge in the connectivity graph and has an associated qubit map. For example, in state 1, gate $g_0$ is assigned to the edge $(p_0, p_1)$ while gate $g_1$ is assigned to the edge $(p_2, p_3)$.
The \swap operations inserted between states dictate how the qubit map changes.

Each additional noisy gate increases the probability of error in the execution of the circuit, and two-qubit gates like the \swap gate are 
especially costly, with an error rate that is typically an order of magnitude higher than single qubit gates.
Therefore, the goal is to find a solution which minimizes the number of inserted \swap gates.

\begin{mybox}
  In summary, the \nisqmr problem consists of the following components.
    \begin{itemize}
      \item \textbf{Input:} a circuit and a \nisq \qpu (represented as a connectivity graph).
      \item \textbf{Output:} an execution plan for the circuit on the \qpu as a sequence of states.
      Each state consists of a map from circuit qubits to \qpu physical qubits  and a set of two-qubit gates that are executed.
      \item \textbf{Gate Realization:} the plan associates each gate with a \emph{realization}, the edge along which it is implemented. 
      \item \textbf{Transitions:} between states, the qubit map can be transformed by \swap gates, which define the valid \emph{transitions}.
      \item \textbf{Cost:} the goal is to minimize the number of added \swap operations.
    \end{itemize}
        
\end{mybox}

\subsubsection{Surface Code} 
\begin{figure}
    \begin{subfigure}[b]{0.45\linewidth}
        \centering
        \includegraphics[scale=.7]{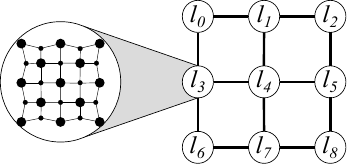}
        \caption{A $3\times 3$ surface code \qpu with 9 logical qubits}
        \label{fig:scmr-gridgraph}
    \end{subfigure}
    \begin{subfigure}[b]{0.45\linewidth}
        \centering
        \includegraphics{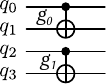}
        \caption{A quantum circuit}
        \label{fig:scmr-circ}
    \end{subfigure}
    \par\bigskip
  \begin{subfigure}{\linewidth}
    \centering
    \begin{tikzpicture}

      \node (step1-table) at (0,0) {
\begin{tabular}{ll}
    \toprule
    Qubit Map & Gates \\
    \midrule
    $q_0 \mapsto l_0$ & \\
    $q_1 \mapsto l_6$ & $g_0 \mapsto  \textcolor{myGreen}{\rule{6pt}{4pt}}$ \\
    $q_2 \mapsto l_2$ & \\
    $q_3 \mapsto l_8$ & \\
    \bottomrule
\end{tabular}
      };
  
      \node[left=.25cm of step1-table] (step1-fig){
        \includegraphics[scale=0.85]{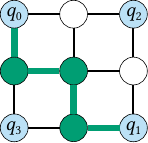}
      };

      \node[above=-0.1cm of step1-table] (step1-label){
        State 1
      };
  
      \node (step2-table) at (4.5,0)   {
          \begin{tabular}{ll}
          \toprule
          Qubit Map & Gates \\
          \midrule
          $q_0 \mapsto l_0$ & \\
          $q_1 \mapsto l_6$ & $g_1 \mapsto \textcolor{myYellow}{\rule{6pt}{4pt}}$ \\
          $q_2 \mapsto l_2$ & \\
          $q_3 \mapsto l_8$ & \\
          \bottomrule
          \end{tabular}
      };

  \node[right=.25cm of step2-table] (step2-fig){
        \includegraphics[scale=0.85]{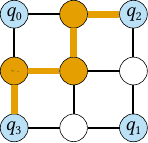}
      };
    \node[above=-0.1cm of step2-table] (step2-label){
        State 2
      };
  
      \draw[->, thick] (step1-table.east) -- node[above]{Id}(step2-table.west);
  
    \end{tikzpicture}
    \caption{A suboptimal \scmr solution that serializes the parallel gates}
    \label{fig:scmr-badmap}
  \end{subfigure}
  \par\bigskip
    \begin{subfigure}{0.48\linewidth}
  \centering
  \begin{tikzpicture}
          \node (table) {

          \begin{tabular}{ll}
          \toprule
          Qubit Map & Gates \\
          \midrule
          $q_0 \mapsto l_0$ & \\
          $q_1 \mapsto l_4$ & $g_0 \mapsto\textcolor{myGreen}{\rule{6pt}{4pt}}$ \\
          $q_2 \mapsto l_1$ & $g_1 \mapsto \textcolor{myYellow}{\rule{6pt}{4pt}}$ \\
          $q_3 \mapsto l_5$ & \\
          \bottomrule
          \end{tabular}
    };
      \node[left=.25cm of table] (fig){
        \includegraphics[scale=0.85]{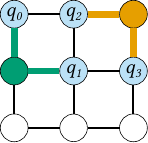}
      };

    \node[above=-0.1cm of table] (label){
        State 1
      };
  
  \end{tikzpicture}
      \caption{An optimal \scmr solution preserving parallelism}
    \label{fig:scmr-goodmap}
\end{subfigure}
~~
    \begin{subfigure}{0.48\linewidth}
        \hspace{.3in}\includegraphics[scale=0.85]{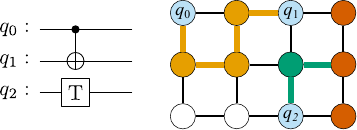}
        \caption{SCMR with a \T gate}
        \label{fig:scmr_with_t}
    \end{subfigure}
    \caption{Overview of the \scmr problem (examples adapted from ~\citet{dascot})}
    \label{fig:scmr-conc}

\end{figure}
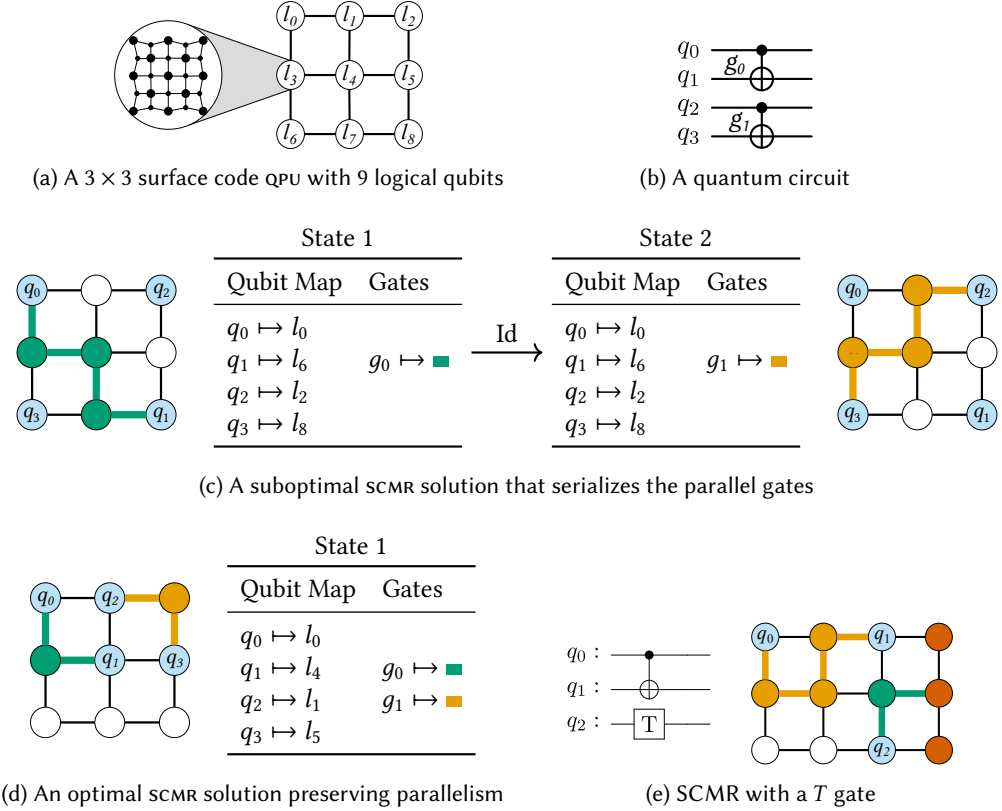

Now we turn to processors implementing the \emph{surface code} \cite{Fowler_2012}, a leading approach for quantum error correction 
that has recently been demonstrated in hardware \cite{Acharya2022SuppressingQE,Google_Quantum_AI_and_Collaborators2024-fp,Bluvstein_2023}. 
We refer to the \qmr problem for surface code processors as \scmr. 
In the surface code, a two-dimensional array of physical qubits is used to encode a fault-tolerant \emph{logical qubit}.
A surface code logical qubit is shown in the inset on the left of \cref{fig:scmr-gridgraph}. The large circles denote physical qubits which carry the logical state,
while the small circles denote physical qubits which are repeatedly measured to detect errors.
A surface code \qpu consists of several surface code logical qubits embedded in the same lattice of physical qubits.
We can represent a \qpu with a grid graph where each vertex represents a logical qubit, and we include an edge between 
adjacent logical qubits, not including diagonals, as shown on the right of \cref{fig:scmr-gridgraph}.

\paragraph{Two-Qubit Gates as Paths}
Two-qubit gates between surface code logical qubits are implemented via a procedure called \emph{lattice surgery} \cite{Horsman_2012_LS}. 
To apply a lattice surgery \cnot gate, we need to find a path of logical qubits on the grid  from the control qubit to the target qubit (through \emph{ancilla qubits}). The path must connect a horizontal boundary of
the control (the top or bottom edge) to a vertical boundary of the target, making at least one ``bend''.
The paths for gates which are executed simultaneously cannot cross. The challenge of \scmr is thus to map circuit qubits to \qpu logical qubits and plan paths such that we avoid conflicts
where one gate blocks another. 

For example, say we wish to execute the circuit in \cref{fig:scmr-circ} on our $3 \times 3$ \qpu in \cref{fig:scmr-gridgraph}.
Theoretically, the two \cnot gates of the circuit can be executed in parallel. 
However, we need to choose the qubit map carefully to enable parallel execution. Consider \cref{fig:scmr-badmap}. With this qubit map,  there is no way to simultaneously execute the two gates,
resulting in a two-time-step solution because any paths from $q_0$ to $q_1$ and $q_2$ to $q_3$ must cross. 
Note that, in this setting, the qubit map does not change between states.
On the other hand, \cref{fig:scmr-goodmap} shows how a different qubit map yields a single-state solution. 
Solutions with fewer states are preferable because they save quantum compute resources at run time, and have a lower probability of logical error, 
which accumulates with each state.

\paragraph{Routing \T Gates}
The surface code problem must also account for \T gates. While the \T gate is a single-qubit gate, it cannot be applied directly to surface code logical qubits. The main proposal for 
addressing this limitation is called magic state injection \cite{Bravyi_2005}. In this protocol, 
a \T gate is implemented via a lattice surgery \cnot gate between the input to the \T gate and another
logical qubit prepared in a so-called ``magic state.'' Magic state qubits are stored in designated
locations on the \qpu, which we represent with special distinguished vertices in our grid graph representation. 

We provide an example in \cref{fig:scmr_with_t}. Since this circuit contains a $T$ gate, we need to define
magic state qubit locations. The $3 \times 3$ architecture has been extended with a column of magic
state qubits along the right side, indicated with orange vertices. An optimal mapping and routing
solution with one state is shown. We have two simultaneous connections: one between
qubits $q_0$ and $q_1$ corresponding to the \cnot; the other between $q_2$ and a magic state
qubit, corresponding to the \T gate.

\begin{mybox}
    In summary, the \scmr problem consists of the following components.
    \begin{itemize}
      \item \textbf{Input:} a circuit and a surface code \qpu (represented as a grid graph). 
      \item \textbf{Output:} an execution plan for the circuit on the \qpu as a sequence of states.
      Each state consists of a map from circuit qubits to \qpu logical qubits and a set of two-qubit gates and \T gates that are executed.
      \item \textbf{Gate Realization:} the plan associates each gate with a \emph{realization}, the path along which it is implemented. 
      \item \textbf{Transitions:} between steps, the qubit map remains constant; that is, the only valid \emph{transition} is the identity transformation.
      \item \textbf{Cost:} the goal is to minimize the number of time steps.
    \end{itemize}
\end{mybox}

\subsubsection{Abstracting the  \qmr Problem}
As these two example problems illustrate, \emph{each \qmr problem can be seen as a specialization of the same abstract \qmr problem}.
Specifically, a \qmr problem specifies a \emph{device state machine} with constraints on the states and transitions between them.
This observation has two major consequences.
First, we can specify a particular \qmr problem succinctly by defining only the unique components of the device state machine: 
\begin{itemize}
\item the problem-specific constraints on valid states (which gates are realizable and how they are realized), 
\item the transition relation 
which describes how the qubit map can change between states, including any additional data that the \qpu graph carries, and
\item the optimization objective defining the cost of a solution.
\end{itemize}
The \lang language is designed to facilitate such a description.  
Second, we can define a generic solver which is parameterized by these components and tries to find an optimal sequence of device states that solves the \qmr problem.
We use the name \framework to refer to either of these components when the meaning is clear from context.

\subsection{The Amaro Language for Specifying \qmr Problems}

To enable painless specification of new \qmr problems,
we design the \framework language, a domain-specific language tailored to our abstract \qmr problem.
In this subsection, we introduce the key features of the \lang language through an example, with a detailed language definition to follow in \cref{sec:lang}.  
An \lang program corresponds to a concrete \qmr problem, like \nisqmr or \scmr.
Through a sequence of data and function definitions written in a simple functional expression language, 
an \lang program defines a function from a circuit and a \qpu (defined as a graph) to a \emph{set} of mapping and routing solutions (as sequences of states), each associated with a cost.
The \solver solver will have to find one of the sequences that minimizes the cost.

For example, the \lang program for the \nisqmr problem is shown in \cref{fig:nisq-qmrl}.
Notice that since \lang is purpose-built for specifying \qmr problems, the programs are concise. 
Here, we fully specify the \nisqmr problem in 12 lines.

\paragraph{Program Structure}
An \lang program takes as input a circuit, referred to implicitly using its constituent gates (\Gate variable), and a \qpu represented as a graph of device qubits (\Arch variable).
The program is divided into a ``RouteInfo'' block (delimited by \programfont{RouteInfo:}) and a ``TransitionInfo'' block (delimited by \programfont{TransitionInfo:}). Together, these two blocks define the device state machine:
\begin{itemize}
  \item
The RouteInfo block specifies the constraints on a valid \qpu state in terms of how a gate is associated with its physical realization.
\item The TransitionInfo block specifies the constraints on a valid transition between \qpu states.
\end{itemize}

\paragraph{RouteInfo Block}
For \nisqmr, a \qpu state associates a \cnot gate with
a pair of \qpu locations.
The first two lines of the \programfont{RouteInfo} block for this problem encode this information. 

The first line defines the \GateRealization datatype. 
In this case, the \GateRealization datatype has a field which is a pair of elements of type \LocT.
The built-in \LocT type represents locations on a \qpu. 
\lang has a built-in notion of an abstract \qpu as a graph over vertices of type \LocT.
The second line specifies that \cnot gates are the gate type to route.

\begin{center}
\begin{lstlisting}
    GateRealization{edge : (Loc, Loc)}
    routed_gates = [CX]
\end{lstlisting}
\end{center}
The rest of the block defines a function \realizegate which says that a \LocT pair is a valid realization for a gate precisely when the pair is an edge between the gate's qubits.
\begin{center}
\begin{lstlisting}
    // realize_gate : (Arch, State, Gate) -> List[GateRealization]
    realize_gate = map(|x| -> GateRealization{edge = x},  
                    Arch.edges_between(State.map[Gate.qubits[0]], State.map[Gate.qubits[1]]))
\end{lstlisting}
\end{center}

The \realizegate function is included in every \lang file. 
It has three implicit parameters, a \qpu \Arch, a state \StateIn, and a gate \Gate.
It returns all possible implementations of the gate which can be added to \StateIn. 
The  \realizegate definition and other mandatory function definitions in \lang omit the fixed parameters from the left-hand side of the assignment for simplicity, and the parameters appear as unbound variables in the expression on the right-hand side.
In this case, \realizegate calls the \GateRealization constructor on each edge in the \qpu \Arch between the 
mapping locations of the two arguments of the \cnot gate in
a functional programming style, using the higher-order function \map.

\paragraph{TransitionInfo Block}
The \programfont{TransitionInfo} block defines the transition relation between
qubit maps of adjacent states. The first line defines the \Transition datatype. In this case, the transitions are 
described by a pair of \qpu locations to which a \swap gate is applied, so once again the datatype has one field which is a pair of elements of type \LocT.
\begin{center}
\begin{lstlisting}
    Transition{edge : (Loc,Loc)}
\end{lstlisting}
\end{center}
 We constrain the pairs to be adjacent in the connectivity graph of the target \qpu with 
the definition of another mandatory function \availtrans.
The function \availtrans has two implicit parameters, a \qpu \Arch and a state \StateIn, and returns a list with elements of type \Transition.
In this case, \availtrans constructs a \Transition for each edge in the \qpu \Arch.
\begin{center}
\begin{lstlisting}
    // get_transitions : (Arch, State) -> List[Transition]
    get_transitions = map(|x| -> Transition{edge = x}, Arch.edges())
\end{lstlisting}
\end{center}

Then, the next line defines a function \apply which describes the action of a transition on a qubit map, which is to swap
the circuit qubits mapped to those locations.
The \lang standard library function \programfont{value\_swap} simplifies this definition.
\begin{center}
\begin{lstlisting}
    // apply : (Transition, QubitMap) -> QubitMap
    apply = value_swap(QubitMap, Trans.edge.(0), Trans.edge.(1))
\end{lstlisting}
\end{center}
Finally, the cost definition says each non-trivial transition (i.e. inserted \swap gate) has a cost of 1. 
\begin{center}
\begin{lstlisting}
    // cost : Transition -> Real
    cost = if Trans == IdTrans then 0.0 else 1.0
\end{lstlisting}
\end{center}

\begin{figure}[t]
  \begin{snugshade}
\begin{lstlisting}
RouteInfo:
  GateRealization{edge : (Loc,Loc)}
  routed_gates = [CX]
  realize_gate = map(|x| -> GateRealization{edge = x},  
                   Arch.edges_between(State.map[Gate.qubits[0]], State.map[Gate.qubits[1]]))

TransitionInfo:
  Transition{edge : (Loc,Loc)}
  get_transitions = map(|x| -> Transition{edge = x}, Arch.edges())
  apply = value_swap(QubitMap, Trans.edge.(0), Trans.edge.(1))
  cost = if Trans == IdTrans
         then 0.0
         else 1.0

  \end{lstlisting}
  \end{snugshade}
  \caption{The \lang definition of the \nisqmr problem}
  \label{fig:nisq-qmrl}
\end{figure}

This simple example demonstrates the general structure of an \lang program: defining data structures and functions
which form an operational description of the constraints of a \qmr problem.  
\lang offers a few other features that were not needed here but do appear in other \qmr problem definitions. These include another optional block for defining the expected labels on the 
\qpu graph edges and vertices (e.g., to say which locations are reserved for storing magic states in \scmr) and built-in utilities for common graph algorithms like path-finding and Steiner tree construction.

\subsection{Generating a \qmr Compiler from an \lang Program} 
Given an \lang program $P$, we can generate a \qmr compiler $\emph{Comp}_P$ for the \qmr problem it specifies  (recall \cref{fig:qmr-overview}).
Specifically, given a circuit and a \qpu graph, $\emph{Comp}_P$ should construct a minimal cost sequence of device states that solves the \qmr problem.
We generate the compiler by \emph{instantiating} the generic \framework \qmr solver for the given \lang program $P$.
The algorithm underlying the \solver solver iteratively constructs a sequence of device states satisfying the definitions in $P$---like the ones in \cref{fig:nisq-conc} and \cref{fig:scmr-conc}---that execute all gates in a given circuit on the given \qpu.

\qmr problems are generally \textsc{np}-hard, and it has been shown that optimal approaches (e.g., using \textsc{sat} solvers) do not scale to large circuits and devices. 
\solver is an approximate algorithm:
In each iteration, \solver constructs a \emph{maximal state}, one in which no more gates can be executed in parallel.
The states depicted in \cref{fig:nisq-conc} and \cref{fig:scmr-conc} are all examples of maximal states.

We identify a key property of \qpus, which we call \emph{monotonicity}, that allows us to construct maximal states without requiring search (all practical \qpus are monotonic).
We also identify the \emph{non-interference} property---executing one gate
does not block another---which allows us to construct \emph{maximum} states without requiring search.
By statically analyzing \lang programs, we can identify non-interference and disable the search process, thus speeding up the solving. For example, \nisqmr is non-interfering while \scmr is not, because the paths of one gate can block another gate's paths.
%

%
As we demonstrate later, (1) we used \solver to generate a diverse range of \qmr compilers for noisy and fault-tolerant \qpus; (2) the generated compilers are competitive with specialized \qmr solvers in terms of solution quality and runtime.

\section{The Abstract \qmr Problem}
\label{sec:prob-def}
In this section, we formally define the abstract \qmr problem. %

\subsection{Circuits and \qpus}
We begin by defining the two inputs to the \qmr problem: circuits and \qpus. We view a circuit
as a sequence of applications of quantum gates. Throughout, we fix a universe of gates \gates that can 
appear in a circuit and a universe of qubits \qubits that they can be applied to.

\begin{definition}[Instruction]
  An \emph{instruction} $\gate(\overline{q})$ is a quantum gate $\gate \in \gates$ acting on a list of qubits $\overline{q}$ drawn from \qubits.
  We denote the set of all instructions as \indexedgates.
\end{definition}

\begin{definition}[Quantum circuit]
A \emph{circuit} is an indexed sequence $g_1(\overline{q}_1), \ldots, g_k(\overline{q}_k)$ of instructions. (We will often drop the $\overline{q}_i$ and refer to the instruction simply as $g_i$.)
\end{definition}

Gates in a circuit which act on the same qubit(s) must be executed in order.
This naturally gives us a partial order on circuit instructions and a notion of a topological layering of a circuit.
\begin{definition}[Gate dependency]
  Instruction $g_j(\overline{q}_j)$ \emph{directly depends} on instruction $g_i(\overline{q}_i)$
  if $i < j$ and some qubit appears in both---that is, $\overline{q}_i \cap \overline{q}_j \neq \varnothing$.
  \emph{Dependency} is the transitive closure of direct dependency and is denoted $g_i < g_j$.
\end{definition}

\begin{definition}[Circuit layer]
  A \emph{layer} is a set of circuit instructions  with no dependencies between them. 
  The \emph{front layer} is the set of instructions that do not depend on any instruction in the circuit.
  \label{def:front-layer}
\end{definition}

\begin{example}
  Consider the circuit in \cref{fig:circ}. The direct dependencies in this circuit---also the only dependencies---are  $g_0 < g_2$, $g_0 < g_3$, $g_1 < g_2$, and $g_1 < g_3$. 
  This circuit has the front layer $\{g_0, g_1\}$.
\end{example}

Next is the representation of a \qpu. 
Our abstraction for a \qpu is a set of locations \locs (typically denoting device qubits) and device instructions \devinstruct.

\begin{example}
    For the \nisqmr problem, we represent a \qpu with a connectivity graph. %
    The set \locs is the graph vertices, which represent physical qubits.
    The device instructions \devinstruct are the edges of the graph. 
    For the \scmr problem, we represent a \qpu with a grid graph. %
    The device instructions \devinstruct are (certain) paths in the graph. 
\end{example}

\subsection{Device State Machines}

To model a time step of execution on a given device, we define the notion of a \qpu \emph{state}.
In essence, a \qpu state is the implementation of a set of gates which can be executed in parallel while satisfying the constraints of the device.
We have seen examples of \qpu states in \cref{sec:examples}. 
As seen in the tabular representations in \cref{fig:nisq-sol,fig:scmr-badmap,fig:scmr-goodmap},
the state of a \qpu consists of two parts. 
One piece of a \qpu state is a qubit \emph{map}, which assigns circuit qubits to locations.
The other is an assignment of instructions in the circuit to device instructions---a set of gate \emph{routes}.
Note that these are both \emph{partial} functions.
The qubit map is only defined on qubits that appear in the circuit, and the gate route is only defined 
on a set of circuit instructions that can be executed in parallel.

\begin{definition}[Device state]
  The state of a \qpu is a pair of injective partial functions:
  \begin{itemize}
    \item $\map : \qubits \rightharpoonup \locs$
    \item $\route : \indexedgates \rightharpoonup \devinstruct$ 
  \end{itemize}
\end{definition}

\begin{definition}[Device state machine]
We call a tuple $(\realizable, \transrel_c)$ a \qpu state machine, where
\begin{itemize}
  \item $\realizable$ is a predicate on device states which determines whether a state is a physically realizable implementation of the circuit instructions in its gate routing.
  \item $\transrel_c$ is a transition relation between states $s \transrel_c s'$ where $c$ denotes the cost of the transition.
\end{itemize}
\end{definition}

\begin{example}
  \label{ex:realizable}
  For the \nisqmr problem, a state is realizable if every two-qubit gate acts on adjacent qubits.
  Let $s = (\map, \route)$ be a state of a \nisqmr \qpu. 
  The predicate $\realizable(s)$ holds if and only if every circuit instruction \cnot $q_i$ $q_j$ in the domain of $\route$ is assigned to the edge $(\map(q_i), \map(q_j))$.

  For \scmr, a state is realizable if gate routing paths are (vertex-)disjoint.
  Let $s = (\map, \route)$ be a state of an \scmr \qpu. 
  The predicate $\realizable(s)$ holds iff the following three conditions are met
  \begin{itemize}
    \item every \cnot instruction \cnot $q_i$ $q_j$ in the domain of $\route$ is assigned to a path from a vertical neighbor of  $\map(q_i)$ to a horizontal neighbor of $\map(q_j)$.
    \item every \T instruction \T $q_i$ in the domain of $\route$ is assigned to a path from a vertical neighbor of  $\map(q_i)$ to a horizontal neighbor of a magic state qubit.
    \item no vertex appears in two distinct paths
  \end{itemize}

\end{example}

\begin{example}
  \label{ex:trans}
For \nisqmr, the transitions are \swap operations between adjacent physical qubits on the \qpu, along with an identity transition which leaves the qubit map unchanged. 
The transition relation $\transrel_c$ relates any pair of states with qubit maps that differ by swapping the locations
of a pair of qubits along an edge of the device
The cost of each nontrivial transition is 1.

For \scmr, the qubit map is fixed throughout, so the only transition is the trivial identity transition. 
We assign a constant transition cost of 1. 
\end{example}

\subsection{Mapping and Routing Solutions}
The goal of \qmr is to find a valid sequence of states that routes all the gates and minimizes cost. 

\begin{definition}[Mapping \& Routing Solution]
  \label{def:abstract-qmr}
  Given a circuit $\circuit$ and \qpu state machine \\ $(\realizable, \transrel_c)$, 
  a \emph{mapping and routing solution} is a sequence
  of states $$s_1 \transrel_{c_1} \cdots \transrel_{c_{k-1}} s_k$$
  such that 
  \begin{itemize}
    \item $\realizable(s_i)$ holds for all $i \in [1,k]$;
    \item each instruction $\gate$ in \circuit appears in exactly one state, which we denote $\mathit{state}(\gate)$; and
    \item if $\gate$ and $\gate'$ are instructions s.t. $ \gate < \gate'$, then $\mathit{state}(\gate)$ appears before $\mathit{state}(\gate')$ in the solution.
  \end{itemize}
\end{definition}

The cost of a mapping and routing solution is simply the sum of the transition costs, $\sum_{i = 1}^{k-1} c_i$. The goal is to find a mapping and routing solution of minimum cost.

\begin{example}
  It is straightforward to verify the solution shown in \cref{fig:nisq-sol} is a mapping and routing solution for the input shown in \cref{fig:nisq-conc} 
  when $\realizable$ and $\transrel_c$ are chosen for the \nisqmr problem as described in \cref{ex:realizable} and \cref{ex:trans}. 
  The cost of this solution is 1. 
\end{example}


\section{A Language for Specifying \qmr Problems}
\label{sec:lang}

In this section, we describe the \lang  language for specifying \qmr problems.

\subsection{\lang Language Design}
The goal of \lang is to define a \qmr problem by specifying a family of similar \qpus with related constraints on their states and transitions.
For example, the \scmr problem is defined by grid graphs of error-corrected qubits (refer to \cref{fig:scmr-gridgraph}),
while the \nisqmr problem is defined by noisy qubits connected by edges (refer to \cref{fig:conn-graph}). 
Notice that we represent a target \qpu with a graph in both cases.
Generally, graphs are the main abstraction of a quantum computer used in formulating \qmr problems.
Consequently, graphs are a core primitive in \lang. 
A program in \lang defines a \qmr problem by describing how to interpret a graph as a \qpu state machine,
filling in the state predicate $\realizable$ and transition relation $\transrel_c$.

\paragraph{Solution Generation} Ultimately, we want to efficiently \emph{generate} solutions for a given circuit and \qpu,
not just verify that a solution is valid.
With this aim in mind, \lang is designed such that $\realizable$ and $\transrel_c$ are defined
``constructively,'' with functions like \realizegate and \availtrans that output lists of valid options,
as opposed to declaratively in terms of constraints.
This design decision enables a natural strategy to find a solution for a given circuit and \qpu.


With the information in an \lang program, we can associate a circuit and graph with a set of valid mapping and routing solutions and determine the cost of each solution.  
In other words, when we write a program $\prog$ in \lang, its semantics is treated as a set of tuples where
\[(\circuit, \graph, \solution) \in \denote{\prog}\]
means that circuit $\circuit$ has a mapping and routing solution $\solution$ (Definition \ref{def:abstract-qmr}) for the \qpu represented by graph~$\graph$.

\subsection{The \lang Grammar}
\begin{figure}
  \begin{minipage}[t]{0.44\textwidth}
    \small
  \[
    \begin{array}{@{}rcl@{}}
      g & \in & \gates \quad\text{(gates)} \\[3pt]
      x & \in & \Vars \quad\text{(variables)} \\[3pt]
      \tau & \Coloneqq & \LocT \alt \Int \alt \Float \alt \Bool \alt \\
      & \alt & \List[\tau] \alt \tau \times \tau \alt \tau \to \tau \alt \StructT \\
      & | & \ArchT \alt \GateT \alt \Qubit \alt  \StateT \\[3pt]
      P & \Coloneqq & \mathit{RBlk}~\mathit{TBlk}~\mathit{ABlk}~\mathit{SBlk} \\[3pt]
      \mathit{RBlk} & \Coloneqq & \begin{array}[t]{@{}l@{}}
          \programfont{RouteInfo:}  \\
          \quad \GateRealization\{\overline{x} \ty \overline{\tau} \};\\
          \quad \begin{array}{@{}l@{\hspace{1ex}}c@{\hspace{1ex}}l@{}}
            \routedgates & = & \overline{g} ; \\
            \realizegate & = & e;
          \end{array}
      \end{array} \\
      \mathit{TBlk} & \Coloneqq & \begin{array}[t]{@{}l@{}}
          \programfont{TransitionInfo:} \\
          \quad \Transition\{\overline{x} \ty \overline{\tau} \};\\
          \quad \begin{array}{@{}l@{\hspace{1ex}}c@{\hspace{1ex}}l@{}}
            \availtrans & = & e ; \\
            \apply & = & e ; \\
            \costFun & = & e
          \end{array}
        \end{array}
      \end{array}
      \]
    \end{minipage}
    \hfill
    \begin{minipage}[t]{0.54\textwidth}
    \small
        \[
    \begin{array}{@{}rcl@{}}
      \mathit{ABlk} & \Coloneqq & \varepsilon \\
      & \alt & \programfont{ArchInfo:}~ \Arch\{\overline{x} \ty \overline{\tau} \} ~;~ \getlocations =  e  \\[3pt]
      \mathit{SBlk} & \Coloneqq & \varepsilon \alt \programfont{StateInfo} : \programfont{cost} = e \\[3pt]
      F & \Coloneqq & \programfont{edges} \alt \programfont{all\_paths} \alt \programfont{steiner\_trees} \alt \\
                                    & \alt & \Push \alt \Map \alt \Fold \alt \Concat \alt \cdots \\[3pt]
      v & \Coloneqq & r \in \RR \alt n \in \NN \alt \mathit{str} \alt \loc(v) \alt (v, v) \\
      & | & \fun{x}{e} \alt S\{\overline{x} = \overline{v}\} \alt [\overline{v}] \\[3pt]
      e & \Coloneqq & x \alt \Arch \alt \Gate \alt \StateIn \alt \Trans \\
      & | & \QubitMap \alt \StateIn.\programfont{map} \alt  \StateIn.\programfont{route}  \\
      & \alt &  \Gate.\programfont{gate\_type} \alt \Gate.\programfont{qubits} \\
      & | & \loc(e) \alt e \otimes e \alt e.x \alt [\overline{e}] \alt e[e] \alt \app{e}{e} \\
      & | & (e, e) \alt \proj{i}{e} \alt \ITE{e}{e}{e} \alt F(\overline{e}) \\
      & | & \GateRealization\{\overline{x} = \overline{e}\} \\
      & | & \programfont{IdTrans} \alt \Transition\{\overline{x} = \overline{e}\}
    \end{array}
  \]
  \end{minipage}
\caption{The grammar of \lang programs} 
\label{fig:grammar}
\end{figure}

The syntax of \lang is presented in \cref{fig:grammar}. Here, $\prog$ is an \lang program, and 
the non-terminals with ``-$\mathit{Blk}$'' names are the definition blocks which comprise an \lang program. There are four of these blocks:
the mandatory \programfont{RouteInfo} and \programfont{TransitionInfo} blocks and the optional \programfont{ArchInfo} and \programfont{StateInfo} blocks.
The \programfont{ArchInfo} block defines any labels on the \qpu graph, such as the locations of magic state qubits in \scmr.
The \programfont{StateInfo} block defines a cost function 
on states, for convenience when assigning a cost to a step is a more natural formulation than a transition cost.\footnote{Any \qmr problem with costs on states can be expressed with transition costs alone, so this block is purely for ease of use.}

Within each block are definition lines, which have a keyword on the left-hand side and one of three options on the right. 
First, there is the \routedgates line, for which the right-hand side is a list of gate identifiers (e.g. ``\cnot'' or ``\T'') drawn from~$\gates$.
Second, there are data definition lines, where the right-hand side is a list of field names (drawn from the set of identifiers~\Vars) and their types, which define the datatype associated with the block.
Finally, there are function definition lines, where the right-hand side is an expression from a standard functional expression language defining a function on fixed implicit arguments.


\subsection{\lang Semantics}
\begin{figure}[t]
\centering
  \begin{mathpar}[\small]
    \Rule{FullProg}{
      \solution = s_1 \to_{c_1}  \ldots \to_{c_{k-1}} s_k \ \\
      \prog,  \graph \vdash \forall i \in [1,k].  \realizable(s_i) \\\\
      \prog,  \graph \vdash \forall i \in [1,k-1]. (s_i \transrel_{c_i} s_{i+1}) \\
      \dr(\circuit, \solution) 
    }{
      (\circuit, \graph, \solution) \in \denote{\prog}
    }
    \and
    \Rule{RealEmpty}{
    \mathit{range}(\map) \subseteq \locs}{\prog, \graph \vdash \realizable((\map, \varnothing))}
    \and
    \Rule{RealIns}{
      \prog, \graph \vdash \realizable((\map, \route)) \\\\
      r \in \denote{\prog.\realizegate}(\graph, (\map, \route), \ins)   \\
    }{
     \prog, \graph \vdash \realizable((\map, \route\lbrack \ins \mapsto r\rbrack))
    }
    \and
    \Rule{Transition}{
      s = (\map, \route) \\
      s' = (\map', \route') \\
      t \in \denote{\prog.\availtrans}(s, \graph) \\
      \map' = \denote{\prog.\apply} (t, \map)\\
      c = \denote{\prog.\costFun}(t)
    }{
     \prog, \graph \vdash s \transrel_c s'
    }
  \end{mathpar}
  \caption{\lang semantics. The premise $\dr(\circuit, \solution)$ is a predicate that checks that every instruction of the circuit is in the 
  step sequence and the dependencies are respected (latter two conditions of Definition \ref{def:abstract-qmr}).
  }
  \label{fig:lang-semantics}
\end{figure}

The semantics of \lang programs are presented in \cref{fig:lang-semantics}.
Together, an \lang program $\prog$ and a labeled graph $\graph$ define a \qpu state machine $(\realizable, \transrel_c)$.
We use the judgement $\prog, \graph \vdash \psi$ to mean the formula $\psi$ holds for the definitions of $\realizable$ and  $\transrel_c$ given by~$\prog$ and~$\graph$. 
The \ruleref{FullProg} rule defining the semantics of a full program just rephrases the definition
of the abstract \qmr problem (Definition \ref{def:abstract-qmr}) for the \qpu state machine defined by the program and graph.

\begin{definition}[\lang semantics]
  The semantics of an \lang program $\prog$, written $\llbracket \prog \rrbracket$, is the set of all tuples satisfying the inference rules in \cref{fig:lang-semantics}.
\end{definition}


The other rules use the semantics of expressions~$\denote{e}$ as functions.
Formally, we equip the \lang expression language with a standard deterministic small-step operational semantics.
For an expression~$e$ with free variables~$\overline{x} = \mathrm{fv}(e)$,
we define~$\denote{e}$ to be the partial function that, on inputs~$\overline{v}$,
evaluates $\subst{e}{\overline{x}}{\overline{v}} \stepsto^* w$ and returns~$w$.
If $\subst{e}{\overline{x}}{\overline{v}}$ gets stuck, then $\denote{e}(\overline{v})$ is undefined.

We also equip \lang with a simple type system to ensure that the supplied code expects the correct arguments and computes the correct types.
Specifically, the expressions in the mandatory function definitions have the following signatures:
\[
  \def\arraystretch{1.1}
  \begin{array}{rcl}
    \prog.\realizegate & : & \ArchT, \StateT, \StateT \to \List[\GateRealT] \\
    \prog.\availtrans & : & \StateT, \ArchT \to \List[\TransT] \\
    \prog.\apply & : & \TransT, \programfont{QubitMap} \to \programfont{QubitMap} ~ \text{(where $\programfont{QubitMap} = \Qubit \to \LocT$)} \\
    \prog.\costFun & : &\TransT \to \Float
  \end{array}
\]
See Appendix \ref{sec:lang-formal-def} for more details on the \lang type system and semantics.

The \ruleref{RealEmpty} and \ruleref{RealIns} rules formalize the notion that a realizable state is the result of inductive application of the \realizegate function.
The set \locs is the vertices of $\graph$ if not overridden by a $\getlocations$ definition.

The \ruleref{Transition} rule relates the transition relation $\transrel_c$ to the \programfont{TransitionInfo} block of the program.
A pair of states $(s, s')$ satisfies the transition relation if the qubit map of $s'$ results from 
applying one of the transitions returned by \availtrans.
The cost of the transition is given by evaluating the cost function from $\prog$
on the transition.

\section{Solving \qmr Problems via Maximal-State Construction}
\label{sec:solver}
In this section, we describe the \framework solver for the \qmr problems defined in the \framework language.
For a given \lang program $\prog$, we instantiate the solver as $\mathrm{\solver}_\prog$.
$\mathrm{\solver}_\prog$ takes a circuit and graph, and returns a valid mapping and routing solution of the form
$$s_1 \transrel_{c_1} \cdots \transrel_{c_{k-1}} s_k$$
\solver iteratively constructs a sequence of states while ensuring that each state is \emph{maximal}, meaning that it cannot be extended to a realizable state that routes any additional instructions.
Since different choices of initial map can yield solutions of differing costs, we
repeat this iterative construction with several different initial maps and return the best solution, as described in \cref{sec:map-trans-select}.

Given the general hardness of solving \qmr problems, \solver is not guaranteed to find an optimal solution.
Indeed, as we shall describe, we employ \emph{simulated annealing}~\cite{kirkpatrick1983simulatedannealing}---a classic randomized search algorithm---in different parts of the algorithm to steer the search towards better solutions,
inspired by prior work in \qmr and other quantum compilation problems \cite{dascot,synthetiq}.
See Appendix \ref{sec:sim-anneal-params} for details on the parameters we choose to instantiate our simulated annealing search.

\subsection{The \framework Solver: A High-Level View} 
The high-level steps of the \framework solver are shown in \cref{alg:solver}.
The algorithm starts with an empty solution, $\solution$.
In each iteration, the algorithm constructs a state to add to the solution, $\solution$.
The key invariant \solver maintains is that each state $s_i$ that is added to $\solution$ is realizable and that every consecutive pair of states $s_i, s_{i+1}$ in $\solution$ is such that $s_i \transrel_{c_i} s_{i+1}$.

Let us walk through the solver step by step:
\begin{description}[leftmargin=1em,itemsep=3pt,font=\mdseries\itshape]
\item[Line 3:]
  \solver begins by constructing an initial qubit map $\amap$;
in principle, this can be a random map.
In \cref{sec:map-trans-select}, we discuss a more sophisticated strategy that converges on a suitable
initial map through repeated applications of \cref{alg:solver}. 

\item[Line 5:]
In each iteration, we get the \emph{front layer} of instructions from the circuit (see Definition \ref{def:front-layer}).
Recall that this is the set of all independent instructions in the prefix of the circuit; being independent, these instructions can soundly be routed in parallel if the device permits.

\item[Line 6:]
Next, the algorithm tries to route as many of the instructions in the front layer as possible.
This is done by a process we call \emph{maximal-state construction}, which we describe in detail in \cref{sec:maximal-states}.
The constructed maximal state is added to $\solution$ and the routed instructions are removed from the circuit.

\item[Line 8:]
The next step is to update the $\map$ for the next state.
This is done by calling the \availtrans function, which returns a set of valid transitions from the current state.
The algorithm then selects one of the transitions and updates the $\map$ accordingly.
We describe a strategy for selecting the next transition in \cref{sec:map-trans-select}.
\end{description}

The process repeats until all instructions in the circuit have been routed.

\begin{example}
As an example, we walk through how this algorithm could be used to construct the \nisqmr solution from \cref{sec:examples} (\cref{fig:nisq-sol}).
We begin by choosing the initial qubit map depicted, which maps $q_i$ to $p_i$. 
Then, we find the front layer of this circuit, which includes two instructions: $\cx ~ q_0~q_1$ and 
$\cx ~ q_2 ~q_3$. Both of these instructions can be routed under our qubit map, and so they are included in the maximal state. 
Routing the entire front layer is the best-case maximal state. 

Next, we must choose from the set of four transitions--a \swap along one of the three edges or the identity. 
Not all choices are equal.
If we choose the identity, we fail to make progress in executing the circuit.
At the next iteration of the loop, the front layer consists of the instructions  $\cx ~ q_1~q_3$ and 
$\cx ~ q_0 ~ q_2$. Neither of these are executable, so the maximal state has an empty gate route. 
On the other hand, if we choose to \swap along the edge $(p_1, p_2)$ as depicted in \cref{fig:nisq-sol}, we obtain another maximum set containing both of the candidate instructions.
At this point, the algorithm terminates as the full circuit has been routed.
\end{example}
\paragraph{Correctness and Termination} \cref{thm:soundness} says that any solution returned by \solver is a valid 
mapping and routing solution. Moreover, \solver terminates with a solution
as long as the input program and graph do not define sets of states closed under the transition relation that cannot make progress in execution, as defined precisely in \cref{thm:termination}. 
This condition rules out pathological examples like a \realizegate function that always returns an empty list. 
All of our case study problem instances satisfy the termination condition. Nevertheless, despite the guarantee of 
eventual termination, \solver sometimes fails to find a solution within fixed resource bounds, as shown in the plots in \cref{sec:eval}.
\begin{theorem}[$\mathrm{\solver}_\prog$ Soundness]
  \label{thm:soundness}
  Let $\solution$ be the solution returned by $\mathrm{\solver}_\prog$ on input $(\graph, \circuit)$. 
  Then, $(\circuit, \graph, \solution) \in \llbracket \prog \rrbracket$.
\end{theorem}

\begin{theorem}[$\mathrm{\solver}_\prog$ Termination]
  \label{thm:termination}
  Let $(\realizable, \transrel_c)$ be the \qpu state machine defined by an \lang program $\prog$ and graph $\graph$.
  Suppose that for any \realizable state $s$ and instruction $\gate$,
  there exists a \realizable state $s'$ reachable from $s$ under $\transrel_c$ such that $\gate$ is in the gate route of $s'$. Then, $\mathrm{\solver}_\prog$
  terminates on input $(\graph, \circuit)$ for any circuit $\circuit$.  
\end{theorem}

\begin{algorithm}[t]
  \captionsetup{font=smaller} 

  \smaller
    \begin{algorithmic}[1]
        \Procedure{$\mathrm{\solver}_\prog$}{graph $\graph$, circuit $\circuit$}
        \State Initialize an empty state sequence $\solution$
        \State {\color{ACMDarkBlue}Create initial map, $\amap$} \Comment \cref{sec:map-trans-select}
        \While{$\circuit \neq \varnothing$} \Comment \emph{note we remove routed instructions from $\circuit$}
        \State Compute \exec, the front layer of instructions in $\circuit$
        \State {\color{ACMDarkBlue}Construct a maximal state $s$ for $\amap$ and \exec } \Comment \cref{sec:maximal-states}
        \State Append $s$ to $\solution$ and remove instructions routed in $s$ from $\circuit$
        \State {\color{ACMDarkBlue}Get the next map  $\amap'$ from the  valid transitions given by $s$, $\graph$ and $\prog$} \Comment \cref{sec:map-trans-select}
        \State Set $\amap \gets \amap'$
        \EndWhile
    \State \Return $\solution$
    \EndProcedure
    \end{algorithmic}
    \caption{$\mathrm{\solver}_\prog$ algorithm for finding mapping and routing solutions}
    \label{alg:solver}
\end{algorithm}
\subsection{Maximal-State Construction}
\label{sec:maximal-states}
We now describe the \solver strategy for constructing maximal states.
In the process, we define two conditions on \qpu state machines: \emph{monotonicity} and \emph{non-interference}.
In \cref{thm:max-state}, we use these definitions to classify the \qpu state machines for which we can efficiently find a maximal state and those for which the maximal state is unique.

First, we precisely define a maximal state.

\begin{definition}[Maximal State]
  Consider the states routing instructions from a circuit layer $E$. A realizable state $s = (\map, \route)$ is \emph{maximal} if there is
  no realizable super-state $s' = (\map, \route')$ such that $\route \subset \route'$, where $\subset$ denotes
  strict inclusion.
\end{definition}

We observe that, for the \qmr problems we consider, we can find a maximal state for a set of parallel instructions
with one pass through the instructions, as shown in \cref{alg:max-state}: the algorithm simply iterates through the instructions,
calls \realizegate for each, and if the result is a non-empty set of gate realizations, chooses one to add to the gate route.
\cref{alg:max-state} is an efficient procedure that always produces maximal states for ``reasonable'' settings like our case study problems, but 
does not find a maximal state for any arbitrary \qpu state machine.\footnote{
In fact, we cannot expect a tractable algorithm for finding maximal states without introducing restrictions on the $\realizable$ predicate.
For a (contrived) example, we could define a family of \qpus for the boolean satisfiability problem. In this family,  each \qpu state machine corresponds to a boolean formula,
and the $\realizable$ predicate interprets a subset of the layer as a variable assignment, returning \true on a state if the routed instructions constitute a satisfying assignment to the formula.
}

\subsubsection{Monotonicity and Maximality}
We define the notion of \emph{monotonic} $\realizable$ predicates as a condition that is strict enough to ensure 
\cref{alg:max-state} finds a maximal state, but permissive enough to include any realistic \qpu state machine.
Indeed, all of our case study problems are monotonic. 

\begin{definition}(Monotonicity)
  A predicate $\realizable$ is \emph{monotonic} if whenever a state $s = (\map, \route)$ is realizable, so is any sub-state $s' = (\map, \route')$ 
with $\route' \subseteq \route$. 
\end{definition}

\begin{algorithm}[t]
  \captionsetup{font=smaller} 

  \smaller
  \begin{algorithmic}
  \Procedure{route-one-pass}{graph $\graph$, program $\prog$, layer $\exec$, qubit map $\amap$}
  \State Initialize the state $s$ with qubit map $\amap$ and empty gate route set \route.
  \For{each instruction $g$ in $\exec$}
    \State Let \textit{route-candidates} = $\llbracket \prog.\text{\lstinline|realize_gate|} \rrbracket(\graph, s, g)$
            \If{\textit{route-candidates} is non-empty}
            \State Set $\route(\textit{ins}) = r$ for some $r \in \textit{route-candidates}$
            \EndIf
  \EndFor       
    \State \Return $s$
    \EndProcedure
    \end{algorithmic}
    \caption{Constructing a maximal state}
    \label{alg:max-state}
\end{algorithm}

For some \qmr problems, we reach the \emph{unique} maximal state with \cref{alg:max-state} regardless of which order we iterate over the layer or which gate realization we select.
We categorize these cases as \emph{non-interfering}. The intuition behind the definition of non-interference
is that iteration order is irrelevant when routing one instruction in a state does not prevent the routing of another. 
The \nisqmr problem is non-interfering, but \scmr is not because the path of one instruction can 
occupy vertices and prevent the routing of another. 

\begin{definition}(Non-interference)
  A predicate $\realizable$ is \emph{non-interfering} if whenever  $s = (\map, \route)$ and 
  $s' = (\map, \route')$ are $\realizable$ states with the same qubit map which route disjoint sets of instructions, the combined state $(\map, \route \cup \route')$
  is also $\realizable$.
  \label{def:non-interference}
\end{definition}
\begin{theorem}
  The procedure \textsc{route-one-pass} produces a realizable state for any \qpu state machine.
  If the predicate $\realizable$ is monotonic, the resulting state is maximal.
  If $\realizable$ is also non-interfering, then the maximal state is unique in terms of routed instructions.
  \label{thm:max-state}
\end{theorem}

If the predicate $\realizable$ is \emph{not} non-interfering, then maximal states can have different sizes.
We search for the iteration order which maximizes the number of routed instructions.
The full maximal step procedure with this case is shown in \cref{alg:max-state-full}.
Notice how the algorithm is identical to \cref{alg:max-state} except for the addition of the search over permutations of the instructions.
We implement this search with simulated annealing.
Each search step of simulated annealing randomly swaps the position of two instructions in the order.
The cost of a candidate permutation is the number of routed instructions in the constructed maximal state.

\begin{algorithm}[t]
  \captionsetup{font=smaller} 

  \smaller
  \begin{algorithmic}
  \Procedure{route}{graph $\graph$, program $\prog$, layer $\exec$, qubit map $\amap$}
  \State Initialize the state $s$ with qubit map $\amap$ and empty gate route set \route.
    \State Let $\Sigma$ be the set of all permutations of \exec
            \State $\sigma^* \gets \argmax_{\sigma \in \Sigma} |\textsc{route-one-pass}(\graph, \prog, \sigma, \amap)|$
            \State \Return $\textsc{route-one-pass}(\graph, \prog, \sigma^*, \amap)$
    \EndProcedure
    \end{algorithmic}
    \caption{Search for the best maximal state}
    \label{alg:max-state-full}
\end{algorithm}

\subsection{Selecting an Initial Map and Transitions}
\label{sec:map-trans-select}
Ultimately, the goal of \solver is to find a low-cost solution.
We now describe how we select the initial qubit map and transitions with the goal of minimizing solution cost.

\paragraph{Initial Map} Our algorithm searches for the initial qubit map that yields the best solution.
Instantiating $\map$ with any 
qubit map will yield some valid mapping and routing solution.
However, different choice of initial qubit maps often lead to solutions of different cost, as we saw with the two 
examples in \cref{fig:scmr-badmap,fig:scmr-goodmap}.  
We explore the space of possible initial qubit maps to find the one which results in the lowest cost mapping and routing solution. 
The search for a minimum cost solution over initial qubit maps is implemented via simulated annealing.
Each search step of simulated annealing modifies the initial qubit map by exchanging the mapping locations of two qubits or moving a qubit into an unused location,
then applies \cref{alg:solver} with this choice of \map. 
The cost of an initial qubit map in this simulated annealing search is the cost of the final mapping and routing solution.

\paragraph{Transitions} Recall that the \solver algorithm maintains a sequence of states $\solution$ which is a valid solution to the \qmr problem.
In each iteration, the algorithm constructs a state to add to the solution, $\solution$, along with a transition to the next state.
There are many ways to choose the next transition; we simply choose a transition $t$ that maximizes the next state's size in terms of number of routed instructions:
\begin{equation}
  \max_{t} |\textsc{route}(\graph, \prog, \exec',\llbracket\prog.\text{\lstinline|apply|} \rrbracket(t, m))| - \llbracket\prog.\text{\lstinline|cost|} \rrbracket(t)
  \label{eq:choose-trans}
\end{equation}
Here, $\exec'$ is the front layer of instructions in the circuit after removing instructions routed in the current, as well as previous, states.

\subsection{Key Optimizations}
\label{sec:optimizations}

We now describe some of the key optimizations that improve the performance of \solver.

\paragraph{Static Analysis for Non-Interference}
By \cref{thm:max-state}, if $\realizable$ is non-interfering, the search over the space of maximal states in \cref{alg:max-state-full} is unnecessary because there is a unique maximal state.
We identify non-inference in practice through a simple static analysis of the \lang program.
We claim that if the \realizegate definition does not contain the subexpression \lstinline|State.route|,
then the resulting $\realizable$ predicate must be non-interfering. 
Notice that a counterexample to non-interference implies the existence of an instruction $g$ and two states $s = (\map, \route)$ and $s' = (\map, \route')$ such that
$\llbracket \prog.\text{\lstinline|realize_gate|} \rrbracket$ is nonempty for $s$ and empty for $s'$.
However, if \realizegate definition does not contain the subexpression \lstinline|State.route|, then it must evaluate to the same 
result regardless of the input gate route.
Therefore, we can soundly over-approximate the property of non-interference by traversing the \textsc{ast} of the \realizegate definition in search of the subexpression \lstinline|State.route|.
If there is no match, we safely assume non-interference.

\paragraph{Transition Selection Optimizations} 
We close with some additional practical considerations in transition selection.
For one, there are some scenarios where no transition enables execution of any instructions.
For example, consider a \nisqmr problem where the front layer \cnot instructions act on qubits that are more than one \swap away from adjacent.
In order to continue to make progress, we choose the transition that minimizes the distance between qubits acted upon by instructions in the front layer.
Second, we weigh routed instructions by \emph{criticality}, following a strategy introduced in prior \qmr work \cite{dascot,Javadi-abhari-17}. The criticality of an instruction is the length of the longest sequence of dependent instructions which begins with $g$.
We prioritize critical instructions to avoid a long ``tail'' of sparse states as the instructions on the critical path are executed in sequence.

\section{Qubit Mapping \& Routing Case Studies}
\label{sec:case-studies}
To demonstrate the practical utility and flexibility of our approach, we present seven case studies that reflect real-world qubit mapping and routing challenges drawn from the literature. 
These examples span different types of quantum hardware and programming models, each with unique constraints.
%
We select problems (described below) which demonstrate that our design supports:
\begin{enumerate}
  \item  \textbf{Diverse hardware}: There are multiple competing implementations of the physical qubit including superconducting circuits, Rydberg neutral atoms, and trapped ions.
  Physical characteristics of underlying hardware impose constraints on mapping and routing.

  \item   \textbf{Devices with \& without error correction}: We are currently at an inflection point for quantum error correction. 
  Devices that are currently accessible to the public do not implement error correction. 
  However, in recent years we have seen early experimental demonstrations of error correction.
  Emerging error-correcting codes bring new constraints that affect how qubits can be moved, measured, and interacted with. 
  Our framework is designed to handle both modes—pre- and post-error correction—enabling developers to experiment with evolving designs and co-optimize across hardware, \textsc{qec} strategies, and applications.

  \item 
  \textbf{Discrete \& continuous cost functions}: In some cases, the appropriate metric of solution quality 
  is discrete, like the number of added \swap gates or the total number of time steps; in others, the cost function is continuous, like the probability of successful computation.
\end{enumerate}  
\subsection{Architectures without Error Correction}

\paragraph{\nisqmr}
The first case study is the \nisqmr problem described in \cref{sec:examples}. 
Two-qubit gates are only allowed between physical qubits which are adjacent in the connectivity graph. 
\swap gates are inserted to transform the mapping. 
\nisqmr is the relevant \qmr problem when targeting hardware with fixed position qubits (like superconducting circuits)
and no error correction.
This case models today's widely accessible superconducting hardware and reflects the default compilation mode in most current quantum toolchains.
The \lang description of \nisqmr is the example shown in \cref{fig:nisq-qmrl}.

\paragraph{Variable-Error \nisq (\nisqve)} On real quantum hardware, not every link is equally reliable.
Error rates between different two-qubit gates can differ by an order of magnitude~\cite{notallqubits}.
When performing mapping and routing, we should prefer to make use of the two-qubit gates with lower error rates.
When we take variation into account, the optimization objective changes to direct maximization of the probability of successful computation, rather 
than minimization of the number of \swap gates

We can easily capture this version of the problem, which we call \nisqve (for variable error) in the abstract \qmr framework. 
The changes to the \lang definition are shown in \cref{fig:nisq-ve-qmrl}.
Lines identical to the original \nisq are elided. 
Note the addition of the nontrivial ArchInfo block, which carries the data of the reliability of each edge.
In order to use it as an additive cost, this reliability is represented as $-\log(p_{succ})$ where $p_{succ}$
is the probability of error-free gate execution.

We use this data to redefine the cost of a \swap by looking up the success rate of the
corresponding edge. 
Likewise, we add a StateInfo block to define the cost of a step as a sum over the success rates of each of
the implemented gates. 

\paragraph{Trapped-Ions (\tiqmr)} Trapped-ions are an alternative candidate hardware platform.  
Each qubit on a trapped-ion \qpu is implemented as an atomic ion which is trapped with an electromagnetic field \cite{ion_trap_arch}.
Two-qubit gates can be performed between any pair of qubits in the same trap. However, each trap is limited to 10s of qubits. 
Therefore, proposed trapped-ion architectures consist of several interconnected traps.
To perform a two-qubit gate between qubits in different traps, we need to use shuttling operations to physically move 
qubits from one trap to another.
The typical cost function for trapped-ion \qmr (\tiqmr) is the total added time for shuttling operations \cite{bach2025efficientcompilationshuttlingtrappedion,qccedsim}.

\paragraph{Reconfigurable Atom Arrays (\raa)} Another competing hardware platform is the neutral atom quantum computer. 
Each qubit is a Rydberg neutral atom which is trapped in a two-dimensional array of optical tweezers \cite{Bluvstein_2023}, enabling programmable qubit layouts.
Two-qubit gates are executable between qubits within a target radius of one another.
For long-range interactions, atomic qubits can be repositioned over the course of computation by modulating the tweezers.
However, the movements are slow and constrained in direction \cite{enola,olsq-dpqa}.
An entire row or column of the qubit array must be shifted in parallel, and the relative positions of rows and columns is fixed.
In reconfigurable atom array \qmr, each movement operation is associated with an error rate, and the cost function is the probability of successful execution.
As in \nisqve, we convert this to an additive cost by assigning each operation a cost of  $-\log(p_{succ})$ where $p_{succ}$
is the probability of error-free execution.

\begin{figure}

\begin{snugshade}
  \begin{lstlisting}
RouteInfo:
    ...
TransitionInfo:
  ...
  cost = if Trans = IdTrans
         then 0.0
         else Arch.edge_cost[Trans.edge.(0)][Trans.edge.(1)]
ArchInfo:
  Arch{edge_cost : List[List[Float]]}
StateInfo:
  cost=fold(0, 
            |acc,x| -> acc+x, 
            map(|x| -> Arch.edge_cost[State.map[x.qubits[0]]][State.map[x.qubits[1]]], 
                State.route))
  \end{lstlisting}
\end{snugshade}
  \caption{The \lang definition of the \nisqve \qmr problem}
  \label{fig:nisq-ve-qmrl}
\end{figure}

\subsection{Architectures With Error Correction}
\paragraph{\scmr} We begin our study of \qmr in the presence of error correction with the \scmr problem from \cref{sec:examples}. 
In this setting, two-qubit gates and \T gates are implemented by routing paths between locations on the architecture.
Simultaneous paths cannot cross (must be vertex-disjoint). The cost function for this problem is the total number of states.

\paragraph{Interleaved Logical Qubits (\ilq)} The flexibility of configuration in space for Rydberg atoms can be used to implement 
an \emph{interleaved} architecture where logical qubits are ``stacked'' on top of one another \cite{mqlss}. 
In routing two-qubit gates, there is a distinction between two-qubit gates applied to qubits in the same stack (intra-stack gates) and two-qubit gates applied to qubits in different stacks (inter-stack gates). 
Inter-stack two qubit gates are implemented via lattice surgery and take time that scales with the number of physical qubits used per logical qubit (i.e. the code distance), whereas the intra-stack gate takes constant time, independent of the size of logical qubit.
Therefore the cost function assigns a cost of 1 to a state with only intra-stack gates and a cost equal to the code distance to other states. 

\paragraph{Multi-Qubit Lattice Surgery (\mqlss)} 
Some quantum computing models, such as Pauli-based computation, require multi-qubit measurements involving arbitrary sets of qubits. 
The lattice surgery procedure used to implement two-qubit gates on surface code logical qubits can be extended to support multi-qubit measurements \cite{Litinski2018AGO,mqlss}.
Multi-qubit measurements are implemented using branching lattice surgery paths, forming tree structures rather than linear routes. 
This case study generalizes the routing model, helping developers prototype and evaluate alternative computational models for fault-tolerant execution.
The cost of a solution is the total number of states.

\section{Implementation and Empirical Evaluation}
\label{sec:eval}
We implemented \framework as a Rust library ($\sim$6500 LoC).
A \qmr problem $\prog$ defined in \lang is translated to Rust and compiled with the library, generating a binary $\mathrm{\solver}_\prog$.\footnote{Alternatively, the \lang definition can be written directly in Rust. 
For cases where defining problem components requires complex computation, this may be preferable.} 
This binary takes a circuit and a \qpu graph as input and produces a mapping and routing solution.
The generated compiler leverages parallelism by default, instantiating one run of \solver search per allotted CPU core and returning the best result.

We aim to answer the following empirical research question for each of our case studies:
\begin{description}
  \item[Q1] How do our generated compilers compare to problem-specific state-of-the-art approaches in terms of solution quality?
  \item[Q2] How long do our generated compilers take to converge to a solution?
\end{description}

We also evaluate the impact of individual algorithm choices in the \solver solver.
\begin{description}
  \item[Q3] How do initial qubit map search and maximal state search contribute to solution quality?
\end{description}

\paragraph{Benchmark circuits} For each case study, we benchmark compilers with the suite of 243 application circuits collected by \citet{dascot}, which subsumes that of \citet{zulehner2018efficient}. 
This suite contains arithmetic circuits derived from the RevLib suite \cite{wille2008revlib}, programs written in the Quipper \cite{green2013quipper} and ScaffoldCC \cite{JavadiAbhari2014} quantum programming languages.
It also includes implementations of major quantum algorithms: Shor's Algorithm \cite{shor1994}, the Quantum Fourier Transform \cite{coppersmith2002qft}, Bernstein-Vazirani \cite{qcc}, QAOA \cite{farhi2014qaoa}, and Grover's Algorithm \cite{grover1996}.
These circuits range in size from a few qubits and gates to hundreds of qubits and tens of thousands of gates (summarized in Appendix \ref{sec:more-plots}).
\paragraph{Global Experimental Setup} Unless otherwise noted, the following experimental conditions
apply to all empirical evaluations. All compiler runs are allotted a 1-hour timeout on 16 cores of an AMD EPYC™ 7763
2.45 GHz Processor and 32GB of RAM accessed via a distributed research cluster. 
We choose this fixed timeout to compare \solver, which continuously produces solutions over the course of a search, to a wide range of algorithms with different runtime characteristics.
For example, the heuristic \nisqmr compiler \qiskit can solve any instance within a minute, while the \tiqmr tool \shaper fails to terminate within the hour for most benchmarks.
The parameters of \solver are set to fixed default values for all case studies (see Appendix \ref{sec:sim-anneal-params}).

\subsection{Noisy-Intermediate Scale Quantum (\nisqmr)}

\paragraph{Experimental setup} In the empirical evaluation for the \nisq problem, we target three different connectivity graphs from real \qpus:
Rigetti Aspen-M \cite{rigettiqpus}, Google Sycamore \cite{sycamorespecsheet}, and IBM Eagle \cite{ibmqpus}. We compare against two state-of-the-art compilers. 
The first, \qiskit \cite{qiskit2024}, is an industrial quantum programming toolkit which applies the \sabre algorithm \cite{sabre}. 
Like \solver, it greedily builds a solution by adding \swap gates according to a heuristic scoring function.
(We note that \qiskit has been in development for close to a decade and is heavily optimized for the \nisqmr problem.)
The other tool, \qsynth \cite{shaik_et_al:LIPIcs.SAT.2024.26}, attempts to find globally optimal solutions. 
It encodes the \qmr problem into a satisfiability problem and makes queries to an external \sat solver.

\begin{figure}[t]

  \begin{subfigure}{0.27\textwidth}
      \includegraphics[width=\linewidth]{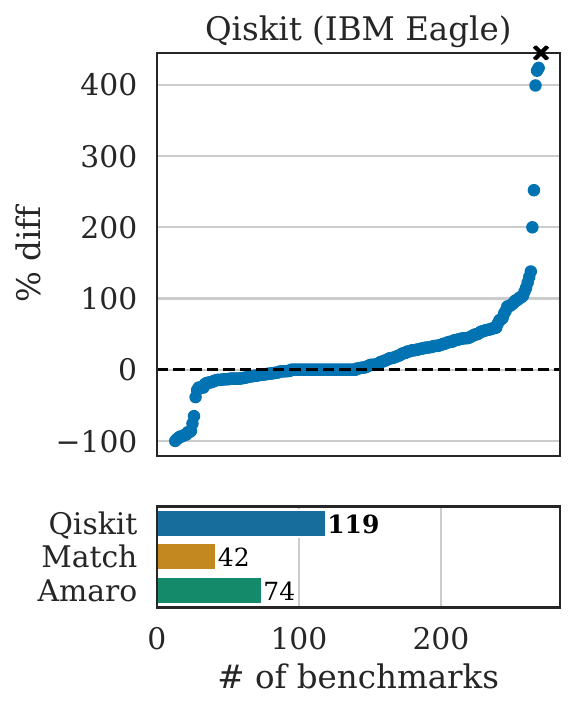}
  \end{subfigure}
  \begin{subfigure}{0.27\textwidth}
      \includegraphics[width=\linewidth]{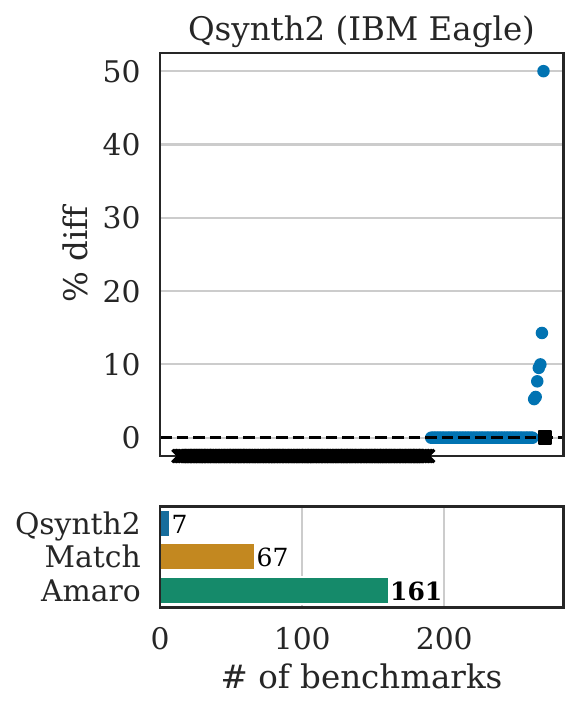}
  \end{subfigure}
  \begin{subfigure}{0.33\linewidth}
  \includegraphics[width=\linewidth]{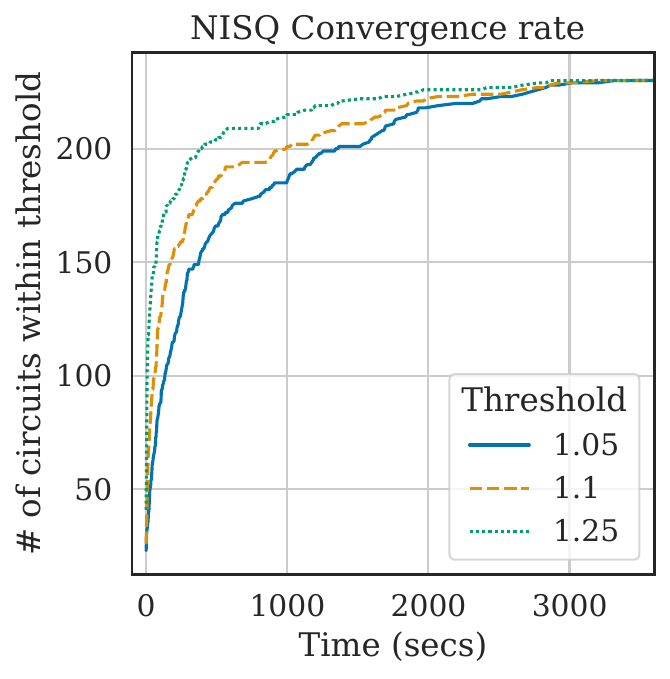}
\end{subfigure}
  \caption{\nisqmr results}
  \label{fig:nisq-results}
\end{figure}

\paragraph{Results} The results are shown in \cref{fig:nisq-results}. In the left and middle plots, we compare the solution quality of our approach to the
baselines on the IBM Eagle architecture (results on other architectures are similar and included in Appendix \ref{sec:more-plots}.
Each point in the scatter plots represents a circuit, and the circuits are sorted by \emph{percent difference} in cost. 
The percent difference is the quantity
\[\frac{\text{ \solver cost} - \text{baseline cost}}{\text{baseline cost}} \]
In this case, the cost is simply the number of added \swap gates.
Points below the dashed line at $y=0$ thus represent circuits where our approach produces a better solution.
Benchmarks where the baseline compiler, \solver, or both fail to find a solution are indicated by black marks on the bottom of the plot (baseline timeout), on the dashed line (both timeout), and on the top of the plot (\solver timeout).
The bar plots shown under the scatter plots aggregate the number of benchmarks below the dashed line of equivalent performance (labeled ``\solver''),  on the line (labeled ``Match''), and above the line (labeled with the baseline), including timeouts.
 
Overall, \solver is competitive with the specialized tools.
For example, in the left plot we see that \solver matches or outperforms \qiskit on roughly 48\% of benchmarks, with \qiskit strictly outperforming \solver on the remaining 52\%, including one timeout.\footnote{8 circuits with too many qubits to be executed on this architecture are excluded from the total} 
Though in the aggregate the two tools have similar performance, there are outliers in both directions where one significantly outperforms the other. The points near and above 400\% difference all correspond to relatively wide and shallow circuits,
with over 100 qubits and under 300 \cnot gates. These are likely workloads where specialized mapping algorithms can find a good initial qubit map among a large search space.

The comparison to \qsynth in \cref{fig:nisq-results}(middle).
\qsynth is optimal, so there are no cases where \solver finds a solution strictly lower in cost. 
\solver solves 161 benchmarks where \qsynth does not terminate.
Excluding timeouts, \solver matches the optimal solution in 66/73 cases.

The line plot on the right of \cref{fig:nisq-results} shows the rate at which \solver converges to a solution. 
We plot the number of circuits for which \solver has found a solution which is close to the final cost of 
the best solution found within the 1hr time out. 
Each line represents a different threshold for ``close.'' The solid blue line is the strictest requirement---current solution is within 5\% of the final solution cost---which is why
it is the lowest line at any particular time point.
In 80\% of the benchmarks, a solution meeting the 10\% threshold is found within 492 seconds.

\subsection{NISQ Variable Error (\nisqve)}

\paragraph{Experimental Setup} To evaluate performance on the \nisqve problem, we augmented the three connectivity
graphs from \nisqmr above with error rates for each coupling link between pairs of qubits. 
These error rates were generated by sampling uniformly from the range $[10^{-3}, 10^{-1}]$, 
to match the scale of two-qubit errors observed on the actual hardware \cite{willowspecsheet,rigettiqpus,ibmqpus}.

\begin{wrapfigure}{r}{0.6\textwidth}
  \centering
  \begin{subfigure}{0.43\linewidth}
    \includegraphics[width=\linewidth]{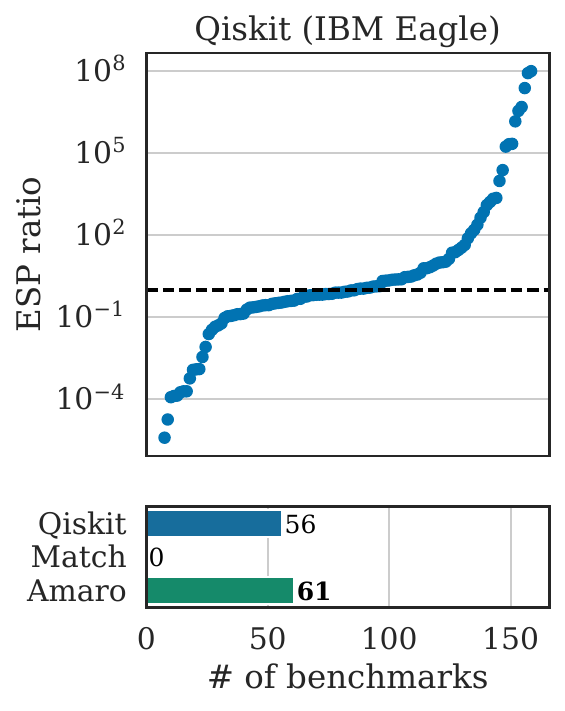}
  \end{subfigure}
  \begin{subfigure}{0.53\linewidth}
    \includegraphics[width=\linewidth]{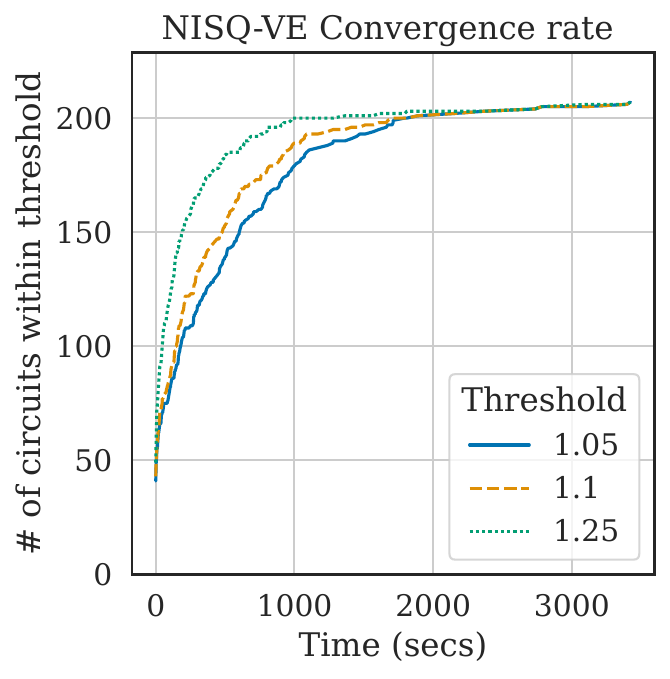}
  \end{subfigure}
  \caption{\nisqve results}
  \label{fig:nisqve-results}
\end{wrapfigure}

\paragraph{Results} The \qiskit compiler can be configured to solve \nisqve. We compare the success probability of \solver to \qiskit in \cref{fig:nisqve-results}(left).
Here, the y-axis represents the \emph{success probability ratio} which is the \qiskit estimated success probability divided by the \solver estimated success probability. Note the logarithmic scale.
Points below the dashed line at $y=1$ represent circuits where our approach produces a better solution.
We exclude benchmarks where one of the computed success probabilities is extremely low, below $10^{-16}$, for numerical stability.
We find that \solver and \qiskit are essentially equivalent on average for the variation-aware variant of the \nisq problem.

Even though \qiskit  produces solutions with fewer \swap gates (as seen in the results for the \nisq problem), this is counterbalanced in the variation-aware setting by the fact that \solver chooses \swap gates with lower error rates.
These results suggest that \solver is an appropriate choice for compilation to \nisq devices in cases where accurate error-rates are readily available.  Moreover, since there are cases where \qiskit significantly outperforms \solver and vice versa,
a portfolio compilation approach may be effective.
The convergence rate of \solver for \nisqve is in \cref{fig:nisqve-results}(right). 
Results are generally similar to \nisqmr: the time to reach the 10\% threshold for 80\% of circuits is 600s.

\subsection{Surface Code Mapping and Routing (\scmr)}

\begin{wrapfigure}{L}{0.22\textwidth}
  \includegraphics[width=\linewidth]{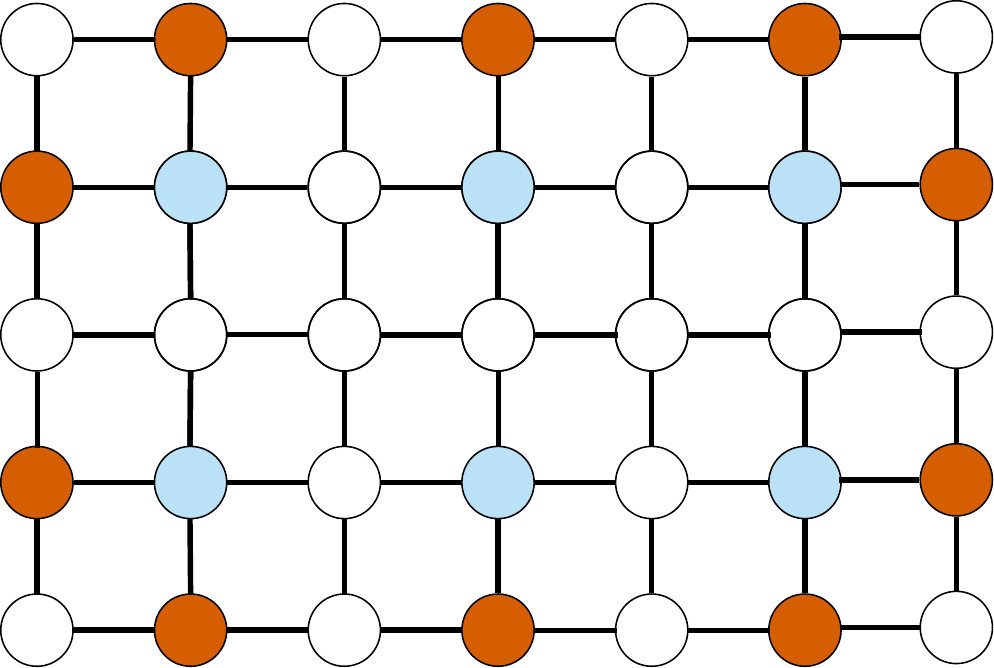}
  \caption{6-qubit \\ Compact Architecture}
  \label{fig:compact-arch}
\end{wrapfigure}
\paragraph{Experimental Setup}
For the \scmr problem, we compare our approach against the state-of-the-art tool \dascot \cite{dascot}.
We target the ``Compact Architecture'' used in the original evaluation of \dascot.
We choose the Compact Architecture because it represents a challenging chase for \scmr with limited 
ancilla qubits available for routing. Logical qubits are arranged almost linearly,  with a row of logical qubits,
a central routing row, then another row of logical qubits. 
Magic state qubits are available along the perimeter of the \qpu. 
An instance of the Compact Architecture with 6 qubits is shown in \cref{fig:compact-arch} (blue and orange vertices denote map locations and magic states respectively).
All circuits are compiled for the smallest possible Compact Architecture \qpu matching the experimental setup originally used to evaluate \dascot \cite{dascot}.

\begin{figure}
\begin{minipage}[t]{0.495\linewidth}
\begin{subfigure}[b]{0.44\linewidth}
  \includegraphics[width=\linewidth]{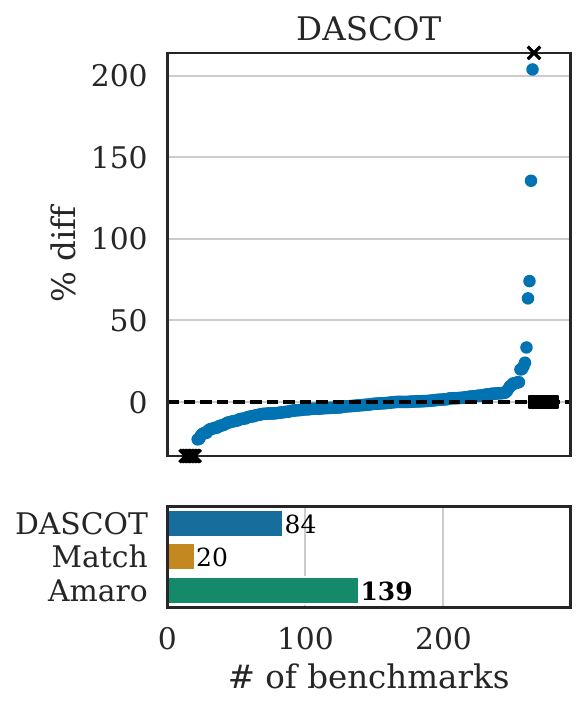}
\end{subfigure}
\begin{subfigure}[b]{0.52\linewidth}
  \includegraphics[width=\linewidth]{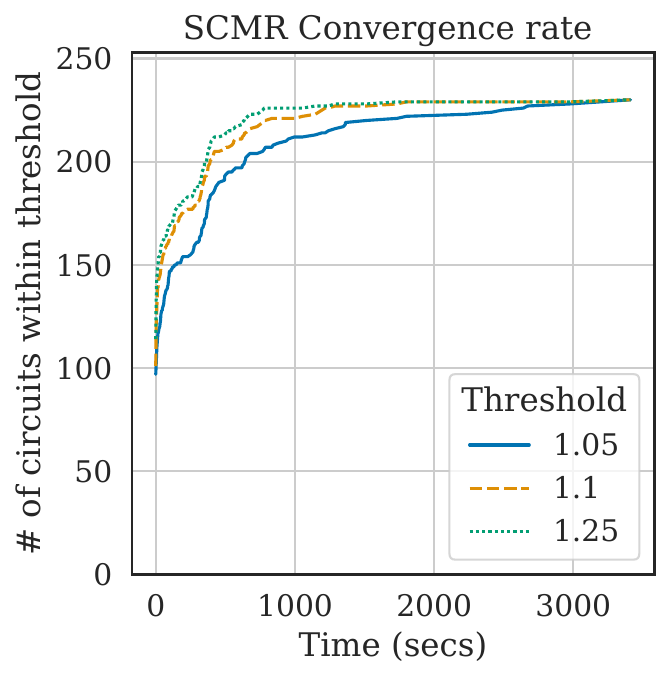}
\end{subfigure}
\caption{\scmr results} 
\label{fig:scmr-results}
\end{minipage}
\begin{minipage}[t]{0.495\linewidth}
\begin{subfigure}[b]{0.42\linewidth}
  \includegraphics[width=\linewidth]{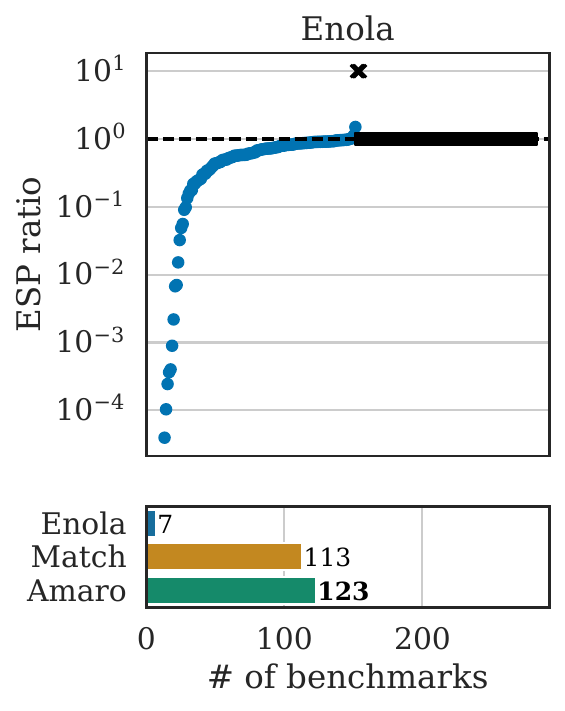}
\end{subfigure}
\begin{subfigure}[b]{0.53\linewidth}
  \includegraphics[width=\linewidth]{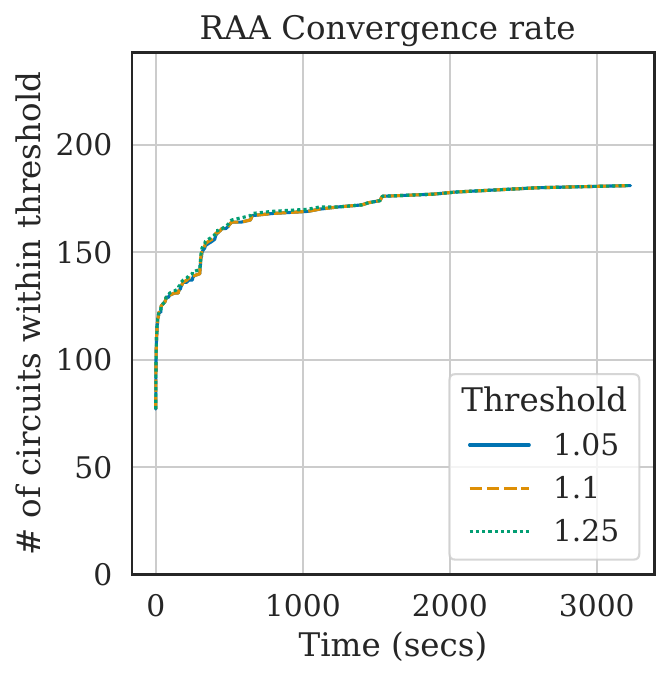}
\end{subfigure}
\vspace*{\fill}
\caption{\raa results}
\label{fig:raa-results}
\end{minipage}
\end{figure}

\paragraph{Results}
\cref{fig:scmr-results}(left) shows a comparison to \dascot in terms of solution quality---number of states in a solution. 
Overall, \solver outperforms \dascot: 
\dascot finds a better solution for 35\% of cases, \solver for 57\%, and the solutions have the exact 
same number of cost or both tools timeout in the remaining 8\%. Solutions  
are often close in quality when both tools terminate, within 6\% for half of the benchmarks.
The two circuits with percent difference over 100\% are both large Grover's algorithm circuits. 
\solver struggles on this relatively large application compared to \dascot.
In \cref{fig:scmr-results}(right), we plot the convergence rate for the \scmr problem. 
Notice that compared to the previous two problems, the first plotted point is higher on the $y$-axis. 
That is, it is more likely that early candidate solutions will be strong choices that are difficult to improve upon.
The time to reach the 10\% threshold on 80\% of the circuits is slightly lower at 324 seconds.

\subsection{Reconfigurable Atom Arrays (\raa)}

\paragraph{Experimental setup} We compare against the state-of-the-art compiler \enola \cite{enola}.  
We use the same values for the empirically derived fidelity parameters (atom transfer fidelity, etc.) as in the original evaluation of \enola. 

\paragraph{Results}
 \solver is able to find slightly higher quality solutions in almost all cases. 
\cref{fig:raa-results}(left) is the same type of plot as \cref{fig:nisqve-results}(left). 
The $y$-axis is the success probability ratio---the \enola estimated success probability divided by the \solver estimated success probability. 
\solver produces a better solution in about 97\% of cases where at least one tool terminates, with a median percent difference of 75\%. 
In the convergence plot, \cref{fig:raa-results}(left), we see that the lines for the three thresholds 
are completely overlapping. The overlap suggests that all benchmarks have a clear best solution found by \solver, and no alternative within 25\% of its cost.
The time to reach the 10\% threshold  on 80\% of the circuits is 304 seconds.

\subsection{Multi-qubit Lattice Surgery (\mqlss)}
\paragraph{Experimental setup} To evaluate 
on the multi-qubit lattice surgery problem, we converted all benchmark circuits to multi-body Pauli product
rotation form \cite{Litinski2018AGO}. The baseline approach presented in \citet{mqlss} is not available for reuse, so we 
compare against a theoretical lower-bound. Any solution for a circuit
with depth $d$ must have at least $d$ states to respect the logical dependencies.

\paragraph{Results}
\solver is able to reach the theoretical lower-bound for most benchmarks within the timeout, though no solution is found for 13. 
Except for these hard instances, \solver converges quickly for this problem, all best solutions (excluding timeouts) are found within the first 10 minutes or so of the search. 
The time to reach the 10\% threshold  on 80\% of the circuits is 44 seconds.

\subsection{Interleaved Logical Qubits (\ilq)}
\paragraph{Experimental Setup} For this problem, we compared against the simple compilation workflow 
used to evaluate the viability of the \ilq architecture as compared to traditional surface code designs \cite{ilq}.
This baseline compiler applies a simple greedy algorithm which tries to group interacting qubits into the same stack, 
then routes instructions as soon as possible.
We target the interleaved version of the Compact Architecture with a stack depth of 4, chosen based on the 
observation by \citet{ilq} that this value is an inflection point, with diminishing returns for large stack sizes.

\paragraph{Results}
\solver outperforms the \ilq baseline on the majority of benchmarks as shown in \cref{fig:ilq-results}(left).
As a greedy algorithm, the baseline is able to solve more benchmarks.
However, \solver finds a better solution in virtually all cases where it terminates, such that 76\% of circuits fall in the \solver better category.
The convergence data for the \ilq problem is shown in \cref{fig:ilq-results}(right).
Results are similar to \raa, with no distinction between the three thresholds 
and requiring 303 seconds to reach the 10\% threshold on 80\% of benchmarks.

\subsection{Trapped-ions (\tiqmr)}

\begin{figure}
\begin{minipage}[t]{0.495\linewidth}
\begin{subfigure}[b]{0.44\linewidth}
  \includegraphics[width=\linewidth]{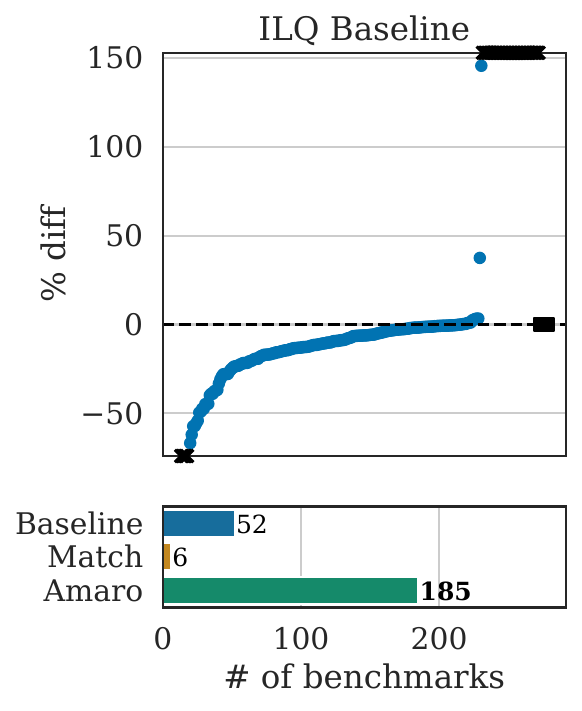}
\end{subfigure}
\begin{subfigure}[b]{0.53\linewidth}
  \includegraphics[width=\linewidth]{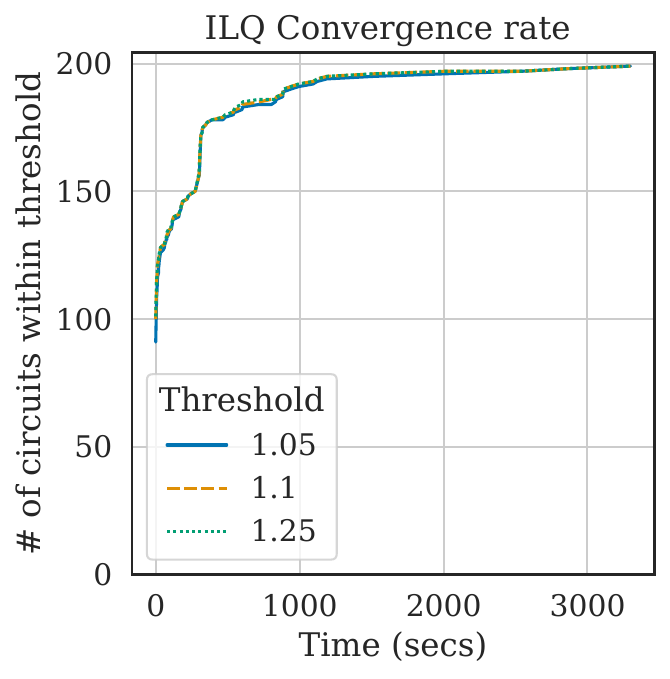}
\end{subfigure}
\caption{\ilq results} 
\label{fig:ilq-results}
\end{minipage}
\begin{minipage}[t]{0.495\linewidth}
\begin{subfigure}[b]{0.44\linewidth}
  \includegraphics[width=\linewidth]{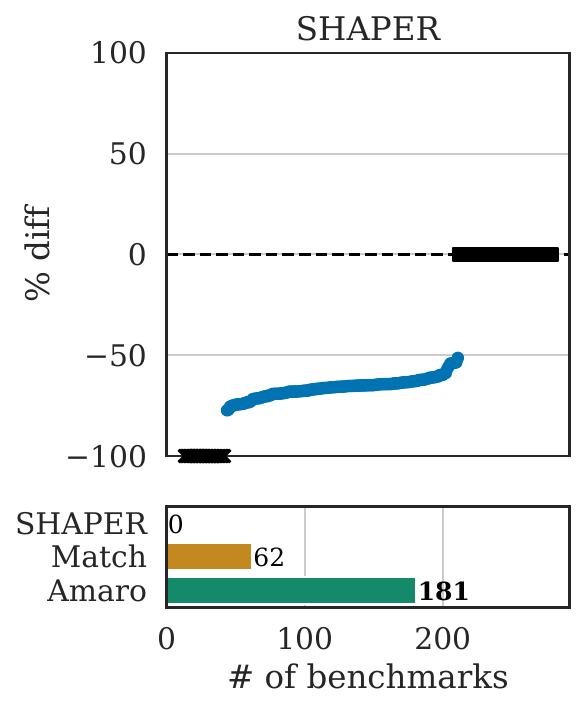}
\end{subfigure}
\begin{subfigure}[b]{0.53\linewidth}
  \includegraphics[width=\linewidth]{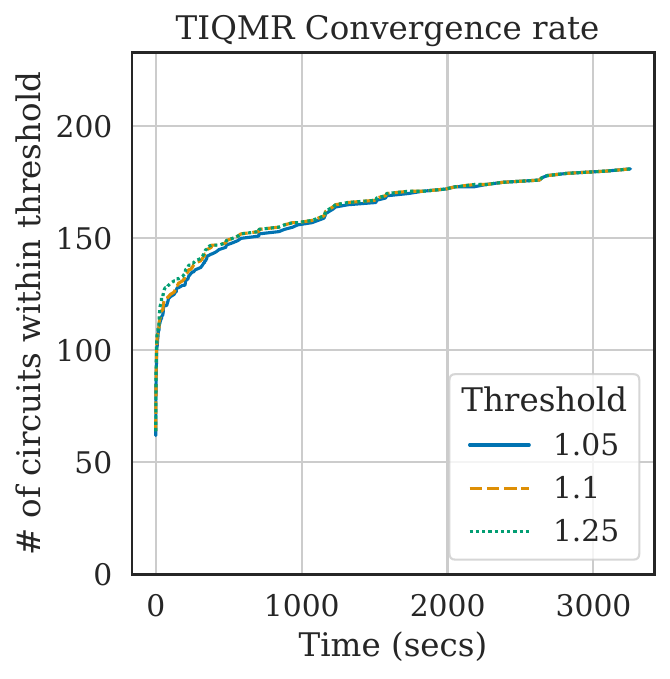}
\end{subfigure}
\caption{\tiqmr results}
\label{fig:tiqmr-results}
\end{minipage}
\end{figure}
\paragraph{Experimental Setup} 
For the trapped-ion problem, we compared against the \shaper algorithm \cite{bach2025efficientcompilationshuttlingtrappedion}.
We target the G2x3 architecture \cite{qccedsim, g2x3}, which consists of 6 traps arranged in two rows with three traps each.
For each circuit, we assume the smallest trap size with capacity for all of the circuit qubits ($\lceil n / 6\rceil$ where $n$ is the number of circuit qubits).

\paragraph{Results}
As shown in \cref{fig:tiqmr-results}(left), \solver significantly outperforms \shaper, always reducing the cost by at least 50\% when both tools terminate. 
It also finds solutions to 27 benchmarks where \shaper does not.
The \tiqmr problem is non-interfering, so here the search over maximal states is not the source of the benefits.
Instead, the results are sensitive to the choice of initial qubit map, and \solver searches over this space more comprehensively than \shaper.
The convergence results for \tiqmr, shown in \cref{fig:tiqmr-results}(right) are similar to other problems,
with a similar time of 352 seconds to reach the 10\% threshold on 80\% of circuits. 
However, we note a less pronounced plateau at the end of the timeout. 
This suggests a longer tail of hard instances for this problem.

\subsection{Ablation Study}
\begin{figure}
\begin{subfigure}{0.25\linewidth}
  \includegraphics[width=\linewidth]{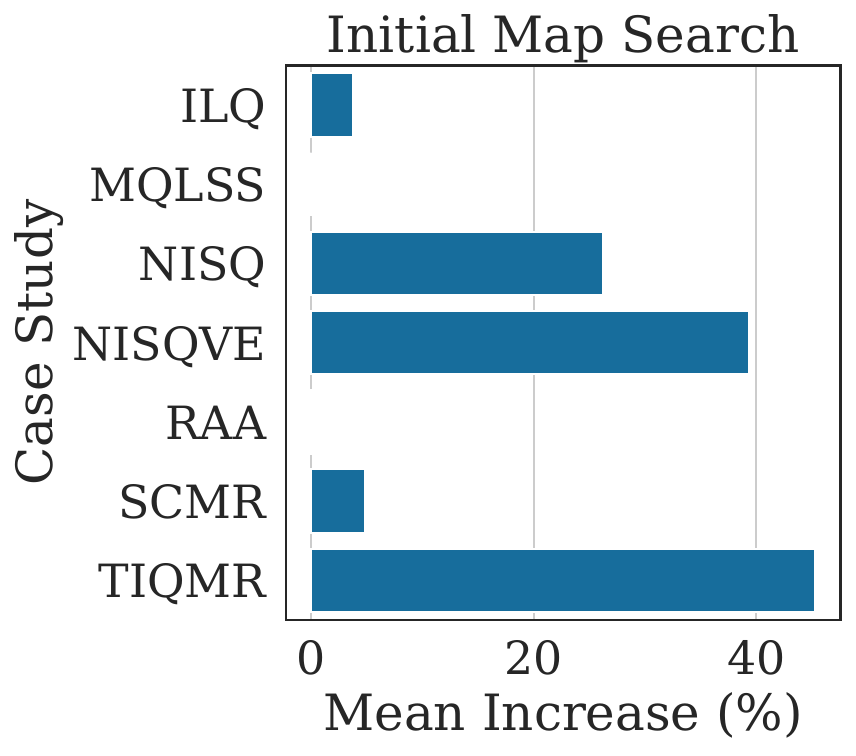}
\end{subfigure}
\begin{subfigure}{0.25\linewidth}
  \includegraphics[width=\linewidth]{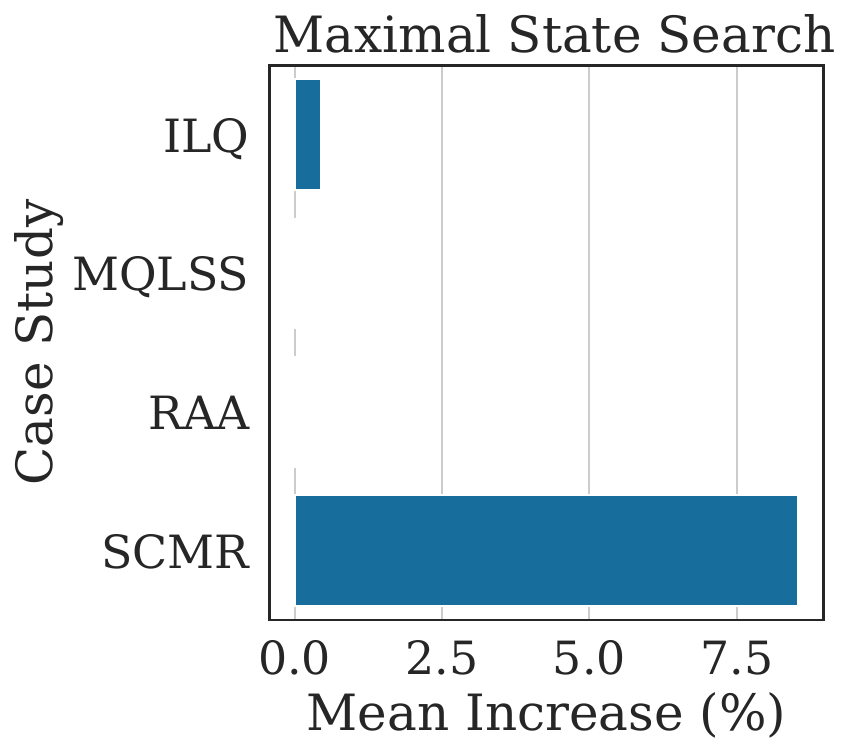}
\end{subfigure}
\caption{Ablation study}
\label{fig:ablation}
\end{figure}

To address Q3, we isolate the effect of the two search subroutines of \solver with an ablation study, with the results shown in \cref{fig:ablation}.

We evaluate the initial qubit map search by comparing to a version of \solver that chooses a single initial qubit map,
rather than applying simulated annealing search to find the best initial qubit map. 
\cref{fig:ablation}(left) shows the percent increase in cost from this ablation for each case study, averaged across the entire benchmark suite (for detailed results, see Appendix \ref{sec:more-plots}).
Here we see some improvement in quality attributable to the initial map search in all but two case studies (on which it has essentially no effect),
including over 25\% improvement in both variants of the \nisq problem. 

We also assess the gains from searching for the best maximal state. To this end, we evaluate against a version of \solver which iterates over a layer in a random order, rather than searching for the best order. 
The results are in the right plot; case studies with non-interferences are excluded because maximal state search is disabled.
In this case, we see a more modest effect, with the \scmr case benefiting the most from maximal state search, at about an 8\% average difference.

\section{Related Work}
\paragraph{Qubit mapping and routing} A large body of prior work has extensively explored \qmr as a manually engineered compiler pass~\cite{QMR_Survey}, and we have discussed representative \qmr algorithms throughout this paper (see \cref{tab:qmr-problems}). %
Our solver shares some structure with the \sabre algorithm for \nisqmr \cite{sabre} which also constructs a single time step at a time, choosing the best transition at each stage.
We also borrow algorithmic insights from \dascot \cite{dascot}, which applies simulated annealing to \scmr. 
However, \dascot uses a two-phase search where the initial map is scored relative to a heuristic function, as opposed to our joint optimization which explicitly computes a solution for each candidate.
Existing algorithms are typically specialized to the constraints of a particular \qmr problem;
to our knowledge, this work is the first to automate synthesis of the \qmr compiler pass. 

\paragraph{Compiler Synthesis}
We are inspired by efforts to automatically synthesize compilers in other domains. %
A classical example is parser generators \cite{BellLabs-TR32:J75,BellLabs-TR39:L75,antlr}, which automatically produce a parser from a grammar. %
Another line of work in this vein is automated program optimization via rewrite-rule synthesis.
Rather than relying on hand-crafted optimizations, rewrite rule synthesizers automatically generate sound substitutions.
Rewrite rule synthesis has been applied to many domains~\cite{herbie,haliderewriting,taso}.
Within quantum computing, rewrite-rule synthesizers have also been developed for quantum-circuit optimization \cite{queso,quartz},
a compiler pass that is typically distinct from \qmr.

\paragraph{Combinatorial search in compilation} \qmr is a compilation pass that requires a combinatorial search 
to find the best solution satisfying constraints of the target hardware. In this way it is related to other compiler 
problems like superoptimization \cite{schkufza2013stochastic} and \textsc{fpga/vlsi} routing~\cite{fpgaarch}.
\qmr even relies on some of the same subroutines as \textsc{vlsi} routing including path-finding and Steiner tree construction \cite{vlsirouting,wireRouting}.

\paragraph{System-modeling DSLs} 
Finally, our use of a domain-specific language for describing the constraints of quantum hardware builds 
on a history of domain-specific languages for modeling systems. 
Examples from software verification include the Spin/Promela model checking framework \cite{concisepromela} for concurrent systems and the Alloy Analyzer \cite{10.1145/3338843}.
In electronic design analysis, hardware description languages like Verilog \cite{10458102} and \textsc{vhdl} \cite{8938196} describe the structure of classical hardware just as \lang describes the structure of quantum hardware.
\section{Conclusion}
There are numerous parallel attempts at building quantum computers using a dizzying array of qubit hardware, physical layouts, and error-correction schemes.
Each combination requires a carefully constructed compiler to map quantum programs onto the quantum processor while satisfying its idiosyncratic constraints.
We have presented an approach for automatically constructing a mapping and routing compiler for a given quantum processor.
We started from the observation that all mapping and routing problems share a similar structure, which we can define using a simple domain-specific language.
Using a generic solving algorithm, we demonstrated that we can construct powerful mapping and routing compilers for a wide range of quantum processors.
We see two avenues for future research:
 (1) Improving the runtime and accuracy of the search algorithm, perhaps using reinforcement learning to construct mapping and routing policies that can transfer between \qpus.
 (2) Combining mapping and routing with circuit-optimization synthesis~\cite{queso, quartz}.
 This allows us to generate compilers that co-optimize the circuit and its mapping onto the device.

\section*{Acknowledgments}
We thank the anonymous reviewers for their insightful feedback and suggestions.
Thanks to Robert Rand, Moin Qureshi, and members of the madPL group for their feedback on early versions of this work.
We are grateful to the CHTC team for maintaining the computing resources which enabled our empirical evaluation \cite{chtc}.
This work is supported by \textsc{nsf} grants \#2340267 and \#2212232 and awards from Meta and Amazon. 

\bibliographystyle{ACM-Reference-Format}
\bibliography{references}
\newpage
\appendix
\section{Proofs}
\paragraph{\cref{thm:max-state}} 
It follows directly from the definition the $\realizable$ predicate for a \lang program (the \ruleref{RealEmpty} and \ruleref{RealIns} rules in \cref{fig:lang-semantics}) that \textsc{route-one-pass} produces a state satisfying $\realizable$. 

Now assume that $\realizable$ is monotonic and let $s = (\map, \route)$ be a state produced by the \textsc{route-one-pass}.
Suppose, for the sake of contradiction that $s$ is not maximal, so there exists some realizable state $s' = (\map, \route')$
where the domain of $\route$ is a strict subset of the domain of $\route'$.
Let $g$ be an instruction in the domain of $\route'$ but not $\route.$ 
Then, there is some intermediate state $s'' = (\map, \route'')$  with $dom(\route'') \subseteq dom(\route) \cup \{g\}$ which is not realizable corresponding to 
the loop iteration where $g$ was visited but not added to $s$.  But then $s''$ is a sub-state of $s'$ with $\route'' \subseteq \route'$,
violating the assumption of monotonicity.

Now assume that $\realizable$ is monotonic and non-interfering and let $s = (\map, \route)$ be a state produced by the \textsc{route-one-pass}
and $s = (\map, \route')$ be another maximal state. 
The combined state $(\map, \route\cup \route')$ is realizable by the definition of non-interference. 
Moreover, this state cannot route any instructions not routed in $s$ since $s$ is maximal. 
The same reasoning holds with $s'$ in place of $s$. 
Therefore, $dom(\route) = dom(\route \cup \route') = dom(\route')$.

\paragraph{\cref{thm:soundness}} 
  Let $\solution$ be the solution returned by \solver on input $(\graph, \circuit, \prog)$.
  By \cref{thm:max-state}, every state in $\solution$ is realizable. Since \solver only attempts to 
  route instructions from the front layer at each iteration, $\dr(\circuit, \solution)$ holds. 
  Finally, it follows directly from the \textsc{transition} inference rule in \cref{fig:lang-semantics},
  that choosing the next map according to line 8 in \cref{alg:solver} ensures that $s_i \transrel_{c_i} s_{i+1}$ for each pair of consecutive
  states. Therefore, we can apply the \textsc{full-prog} rule to conclude
  $(\circuit, \graph, \solution) \in \llbracket P \rrbracket$ as desired.

\paragraph{\cref{thm:termination}}
  We prove the loop in \cref{alg:max-state} terminates by defining an appropriate termination measure. 
  The idea is that only finitely many states in a row can include no routed instructions thanks to the assumption of reachability in the theorem.
  Let $\mu(s) = (i,j)$ where $i$ is the number of unrouted instructions and $j$ is $M - k$ where $M$ is the total
  number of qubit maps for input $(\graph, \prog)$ and $k$ is the number of iterations since the last instruction was routed. 
  Note that the value of $j$ is always positive as long as no state is visited twice because of the assumption that at least one reachable realizable state routes a gate.
  At each iteration, either an instruction is routed or a transition is taken that visits a new state.
  In either case, the measure decreased under lexicographic comparison. Thus, the loop terminates.

\section{Simulated annealing instantiation}
\label{sec:sim-anneal-params}
\paragraph{Acceptance probability}
We accept a new solution $\snew$ to replace a current solution $\scurr$ according to the standard acceptance probability
\[
  \exp{\left\{-\frac{\cost(\snew) - \cost(\scurr)}{\temp}\right\}}
\] 
where $\temp$ is the current temperature.
\paragraph{Parameters}
In our simulated annealing search, we initialize the temperature to a value $\tinit$, reduce by a cooling rate $r$ at each iteration (i.e. multiply the current temperature by $1-r$),
and terminate the search when we reach a final temperature $\tfinal$, or are interrupted by a timeout. 
For the mapping search, we choose $\tinit = 10$, $r = 10^{-3}$ and $\tfinal = 10^{-5}$.
These values were chosen by a grid search on the \nisqmr and \scmr problems over the range $[1, 10^3]$
for $\tinit$ and $[10^{-5}, 1]$ for $r$ and $\tfinal$. 

For the maximal state search in \cref{alg:max-state-full}, we instantiate the search with the same parameters,
along with parallel reduced searches with $r= 10^{-2}, 10^{-1}, 1-10^{10\log(0.9)}$ and  $1$.
Each of these in turn divides the number of iterations by an order of magnitude.
The reduced searches are designed to maximize the chance that at least one valid solution is found.

\section{Language Definition}
\label{sec:lang-formal-def}
Here we present the full type system and semantics for the \lang language.

\subsection{Type System}
\begin{mathparpagebreakable}[\small]
  \VarRule
  \and
  \ArchRule \and \GateRule \and \StateRule
  \and
  \TransRule \and \MapRule \and \IdTransRule
  \and
  \FloatRule \and \IntRule \and \StringRule
  \and
  \LocRule
  \and
  \AbsRule \and \AppRule
  \and
  \PairRule \and \ProjRule
  \and
  \ListRule \and \ListAccessRule
  \and
  \IfRule
  \\
  \ArithFloatRule \and \ArithIntRule
  \\
  \StructRule \and \StructAccessRule
  \and
  \FunRule
\end{mathparpagebreakable}

\begin{mathparpagebreakable}[\small]
  \Rule{RouteInfoOk}{
    \overline{g} \subseteq \gates \\
    \Arch\ty\ArchT, \StateIn\ty\StateT, \Gate\ty\GateT \proves e : \List[\GateRealT] \\
  }{\gates \proves \left({
      \GateRealization\{\overline{x} \ty \overline{\tau} \};~ 
        \routedgates = \overline{g} ;~ 
        \realizegate = e
  }\right)~\textrm{rt-ok}
  }
  \and
  \Rule{TransInfoOk}{
    \Arch\ty\ArchT, \StateIn\ty\StateT \proves e_1 : \List[\TransT] \\
    \transition\ty\TransT, \StateIn\ty\StateT \proves e_2 : \StateT \\
    \transition\ty\TransT \proves e_3 : \Float
  }{
    \proves \left({
      \Transition\{\overline{x} \ty \overline{\tau} \};~ 
        \availtrans = e_1 ; ~
        \apply = e_2 ; ~
        \costFun = e_3 
    }\right)~\textrm{trans-ok}
  }
  \and
  \Rule{ArchOkEmpty}{ }{\proves \varepsilon~\textrm{arch-ok}}
  \and
  \Rule{ArchOk}{
    \Arch\ty\ArchT \proves e : \List[\LocT]
  }{
    \proves \left(\Arch\{\overline{x}\ty\overline{\tau}\};~ \getlocations = e\right)~\textrm{arch-ok}
  }
  \and
  \Rule{StateOkEmpty}{ }{\proves \varepsilon~\textrm{state-ok}}
  \and
  \Rule{StateOk}{
    \StateIn\ty\StateT \proves e : \Float
  }{
    \proves \left(\costFun = e\right)~\textrm{state-ok}
  }
  \and
  \Rule{ProgOk}{
    \gates \proves P.\programfont{RouteInfo}~\textrm{rt-ok} \\
    \proves P.\programfont{TransitionInfo}~\textrm{trans-ok} \\\\
    \proves P.\programfont{ArchInfo}~\textrm{trans-ok} \\
    \proves P.\programfont{StateInfo}~\textrm{trans-ok} \\
  }{\gates \proves P~\mathrm{ok}}
\end{mathparpagebreakable}

\subsubsection{Library function types}
The auxiliary lookup function~$\mathrm{funtype}$ is defined by the following mappings.
\paragraph{List Functions}
\begin{itemize}
  \item $\Push : \List[\tau], \tau \to \List[\tau]$
  \item $\Concat : \List[\tau], \List[\tau] \to \List[\tau]$
  \item $\programfont{contains} : \List[\tau] \to \Bool$
  \item $\programfont{combinations} : \List[\tau], \Int \to \List[\List[\tau]]$
  \item $\Map : (\tau_1 \to \tau_2), \List[\tau_1] \to \List[\tau_2]$
  \item $\Fold : (\tau_1 \to \tau_2 \to \tau_2), \tau_2, \List[\tau_1] \to \tau_2$
  \item $\programfont{combinations} : \List[\tau], \Int \to \List[\List[\tau]]$
\end{itemize}

\paragraph{Graph Functions}

\begin{itemize}
  \item $\programfont{edges} : \ArchT \to \List[\LocT \times \LocT]$
  \item $\programfont{edges\_between} : \ArchT, \LocT, \LocT \to \List[\LocT \times \LocT]$
  \item $\programfont{all\_paths} : \ArchT, \List[\LocT], \List[\LocT], \List[\LocT] \to \List[\List[\LocT]] $
  \item $\programfont{steiner\_trees} : \ArchT, \List[\LocT], \List[\LocT] \to \List[\List[\LocT]] $

\end{itemize}
\paragraph{Instruction Functions}
\begin{itemize}
  \item $\programfont{qubits} : \GateT \to \Qubit$
  \item $\programfont{gate\_type} : \GateT \to \StringT$
\end{itemize}

\paragraph{Other Utility Functions}
\begin{itemize}
  \item $\programfont{horizontal\_neighbors} : \LocT, \Int \to \List[\LocT]$
  \item $\programfont{vertical\_neighbors} : \LocT, \Int, \Int \to \List[\LocT]$
  \item $\programfont{to\_2d} : \LocT, \Int  \to (\Int, \Int)$
  \item $\programfont{value\_swap} : (\Qubit \to \LocT), \LocT, \LocT \to (\Qubit \to \LocT) $
\end{itemize}

\subsection{Semantics}
The semantics of expressions are given by a small-step operational semantics.
Since the language is deterministic and terminating, we use a denotational shorthand of $\denote{e}$
to be the partial function that takes values~$\overline{v}$ for the free variables~$\overline{x} = \mathrm{fv}(e)$,
evaluates $\subst{e}{\overline{x}}{\overline{v}} \stepsto^* w$ and returns~$w$.
If $\subst{e}{\overline{x}}{\overline{v}}$ gets stuck, then $\denote{e}(\overline{v})$ is undefined.

The small-step operational semantics are defined as follows.
\[
  \begin{array}{rcl}
    E & \Coloneqq & [\cdot] \alt \loc(E) \alt E.x \alt E[e] \alt v[E] \alt E \otimes e \alt v \otimes E \alt (E, e) \alt (v, E) \alt \proj{i}{E} \\
    & | & \ITE{E}{e}{e} \alt [\overline{v}, E, \overline{e}] \alt F(\overline{v}, E, \overline{e}) \alt \app{E}{e} \alt \app{v}{E} \alt S\{\overline{x_v} = \overline{v}, x = E, \overline{x_e} = \overline{e}\}
  \end{array}
\]

\begin{mathparpagebreakable}[\small]
  \ECtxRule
  \and
  \EArithRule \and \EAppRule \and \EProjRule
  \and
  \EIfTRule \and \EIfFRule
  \and
  \EStructAccessRule \and \EListAccessRule
\end{mathparpagebreakable}

Operational semantic rules for utility functions are defined below.
\begin{mathparpagebreakable}[\small]
  \EPushRule \and \EConcatRule
  \and
  \EContainsTRule \and \EContainsFRule
  \and
  \EMapRule
  \and
  \EFoldEmpRule \and \EFoldVRule
  \and 
  \ECombinationsRule
  \and
  \EEdgesRule \and \EEdgesBtwnTRule \and \EEdgesBtwnFRule
  \and 
  \EAllPathsRule \and \ESteinerTreeRule
  \and  
  \EQubitsRule \and \EGateTypeRule
  \and 
  \EHorizERule \and \EHorizLRule \and \EHorizRRule \and \EHorizBRule
  \and   
  \EVertERule \and \EVertAbRule \and \EVertBeRule \and \EVertBRule
  \and 
  \ETwoDRule \and \EValSwapLRule \and \EValSwapRRule \and \EValSwapNRule
  \\
\end{mathparpagebreakable}

\section{Solver Optimization: Incremental Isomorphism}
\begin{figure}[h]
  \includegraphics{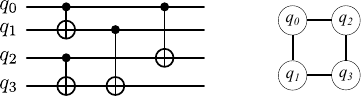}
  \caption{A circuit and corresponding interaction graph}
  \label{fig:circ-with-interact}
\end{figure}

Here we describe another optimization in the implementation of the \solver solver. 
This is a generic optimization which can be applied to all \qmr problems, but it is especially useful for \nisqmr.
As a ``warm-start'' for simulated annealing, we seed the search for a qubit map with a
candidate which places interacting qubits near one another. 
We call this the \emph{incremental isomorphism} optimization because it solves a sequence of subgraph isomorphism problems.
To capture the interactions between qubits in a circuit, we use a well-known data structure called an \emph{interaction graph} \cite{cowtan2019qubit,autobraid, dascot}.
The interaction graph for a circuit includes a vertex for each qubit that appears in the circuit and an edge 
for each pair of qubits to which the circuit applies a two-qubit gate.
An example circuit and its interaction graph are shown in \cref{fig:circ-with-interact}.

\begin{figure}
  \includegraphics[width=0.3\linewidth]{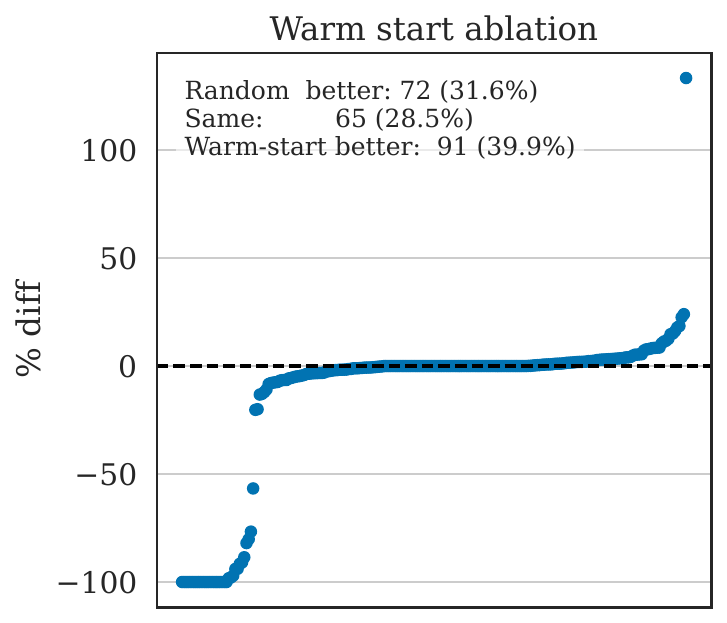}
  \caption{Impact of the incremental isomorphism warm start for \nisqmr}
  \label{fig:inc-isom}
\end{figure}

The incremental isomorphism procedure, shown in \cref{alg:inc-isom}, tracks the interaction graph as it iterates through the \circuit.
Each time an instruction adds a new edge, we check if the current interaction graph can be embedded into the device graph $\graph$.
If so, we set our candidate qubit map according to this embedding. 
Otherwise, we stop iterating and return the current candidate.
The result is a qubit map such that  some prefix of the circuit is likely to be easy to route.
The incremental isomorphism optimization is particularly useful for large circuits with a linear 
interaction graph, like Ising model simulation circuits. 
For these circuits, there is often an embedding of the interaction graph of the full circuit or a long 
prefix which leads to a low-cost solution. 
However, such an embedding is difficult to discover with random search.
See \cref{fig:inc-isom} for empirical results.

\begin{algorithm}[h]
  \captionsetup{font=smaller} 

  \begin{algorithmic}
  \Procedure{incremental-isomorphism}{graph $\graph$, circuit $\circuit$, program $\prog$}
      \State Initialize an empty interaction graph $\interactgraph$ over the qubits in $\circuit$
        \For{each two-qubit instruction in \circuit}
          \State Update $\interactgraph$ with an edge between the qubits of the instruction
          \If{$\graph$ has a subgraph $H$ isomorphic to \interactgraph} \Comment \emph{$H$ need not be an induced subgraph}
          \State Set $\map$ to an isomorphism from $\interactgraph$ to $H$
          \Else
          \State \textbf{break}
          \EndIf

        \EndFor 
    \State \Return $\map$
    \EndProcedure
    \end{algorithmic}
    \caption{Choosing a starting point for the qubit map search} 
    \label{alg:inc-isom}
\end{algorithm}
\section{Additional Plots}
\label{sec:more-plots}
\begin{figure}
  \begin{subfigure}{0.23\linewidth}
  \includegraphics[width=\linewidth]{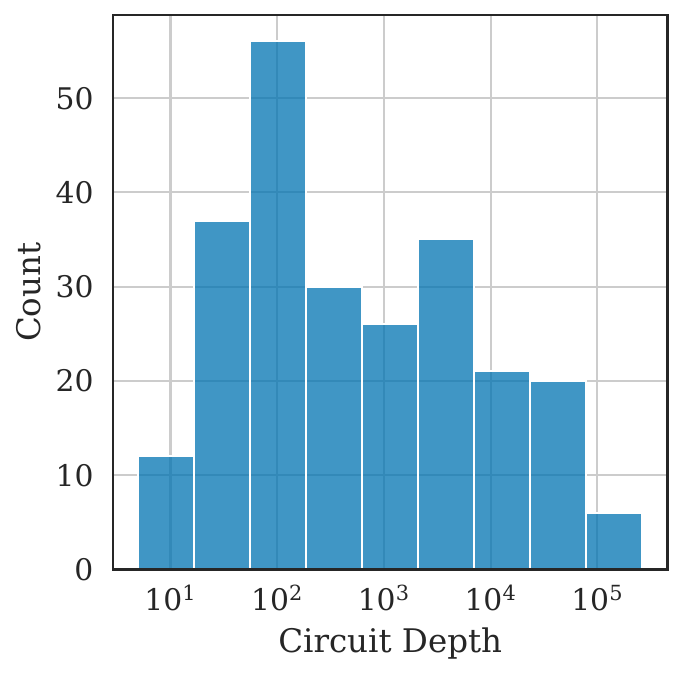}
\end{subfigure}
\begin{subfigure}{0.23\linewidth}
  \includegraphics[width=\linewidth]{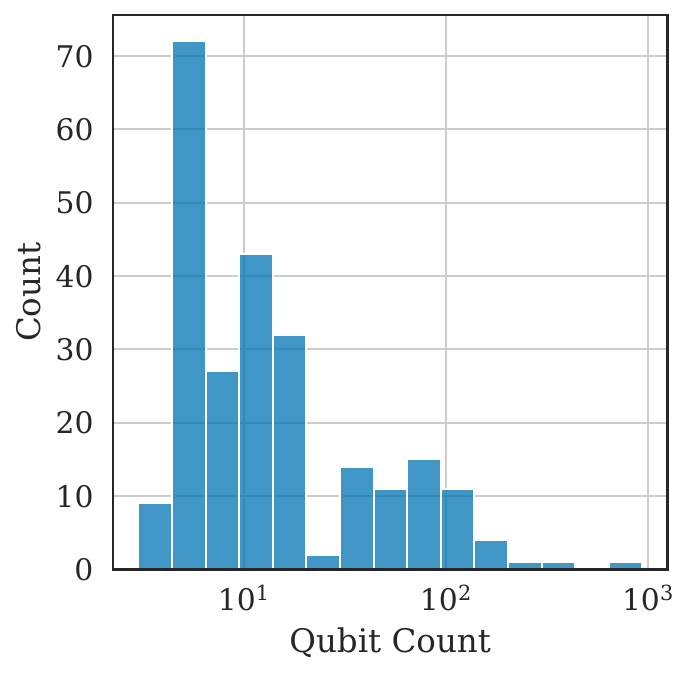}
\end{subfigure}
\caption{Benchmark statistics}
\label{fig:circ-stats}
\end{figure}

\begin{figure}[h]
\begin{subfigure}{0.23\linewidth}
  \includegraphics[width=\linewidth]{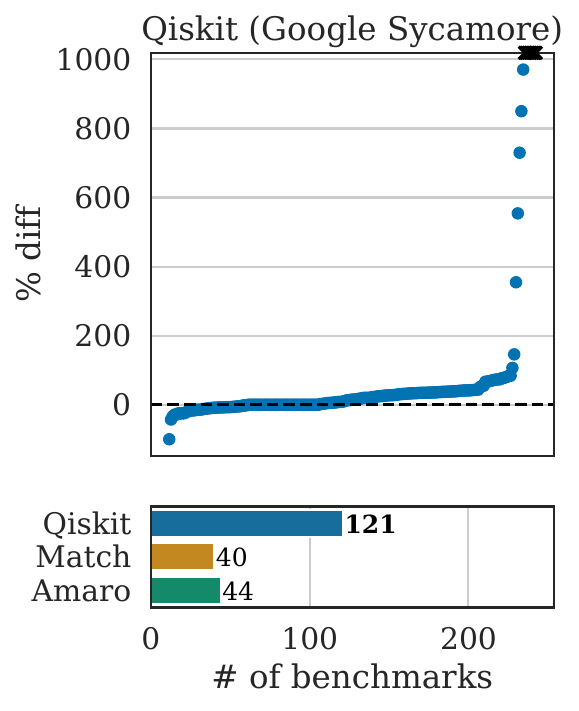}
\end{subfigure}
\begin{subfigure}{0.23\linewidth}
  \includegraphics[width=\linewidth]{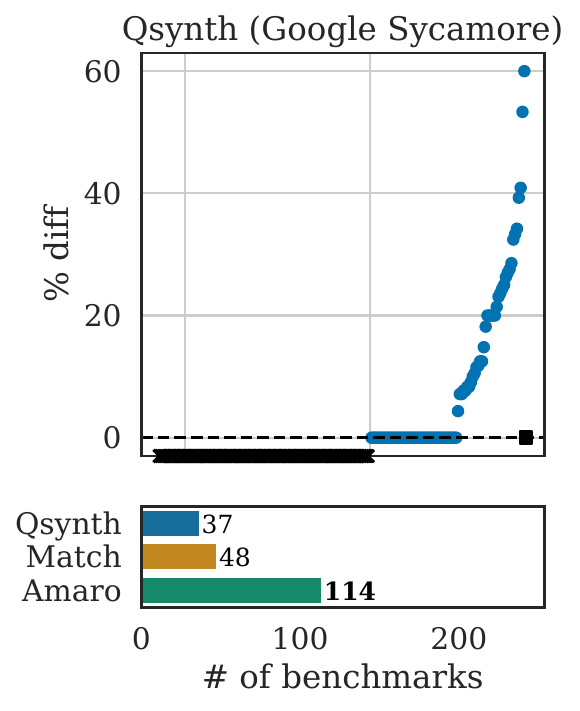}
\end{subfigure}%
\begin{subfigure}{0.23\linewidth}
  \includegraphics[width=\linewidth]{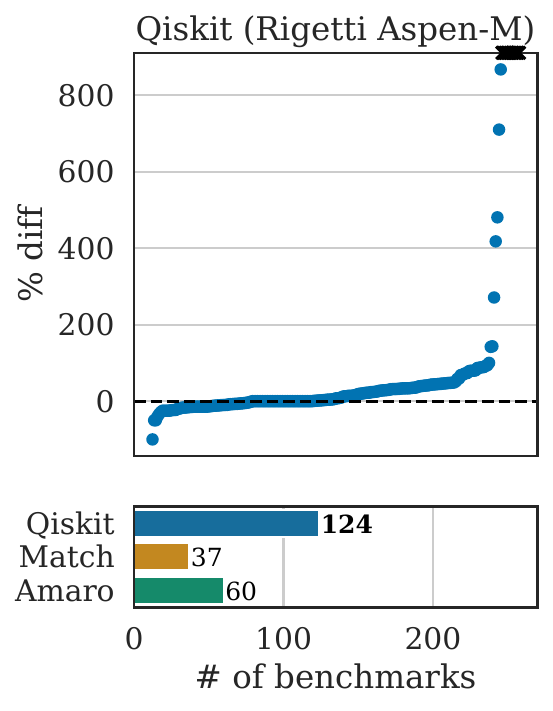}
\end{subfigure}
\begin{subfigure}{0.23\linewidth}
  \includegraphics[width=\linewidth]{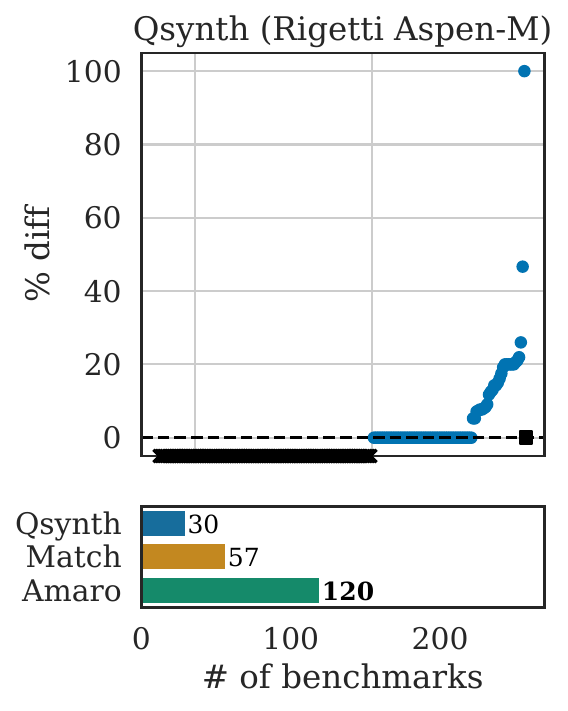}
\end{subfigure}
\caption{\nisqmr results on other architectures}
\label{fig:other-nisq-archs}
\end{figure}

\begin{figure}[h]
\begin{subfigure}{0.23\linewidth}
  \includegraphics[width=\linewidth]{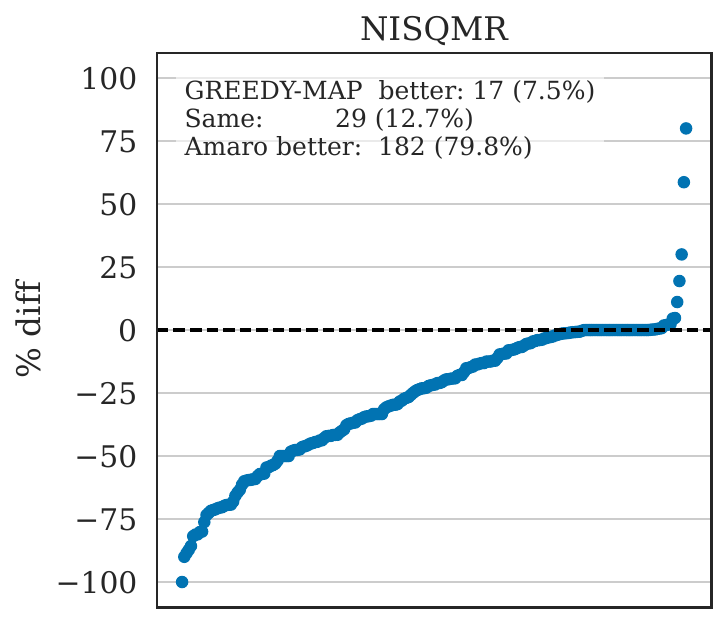}
\end{subfigure}
\begin{subfigure}{0.23\linewidth}
  \includegraphics[width=\linewidth]{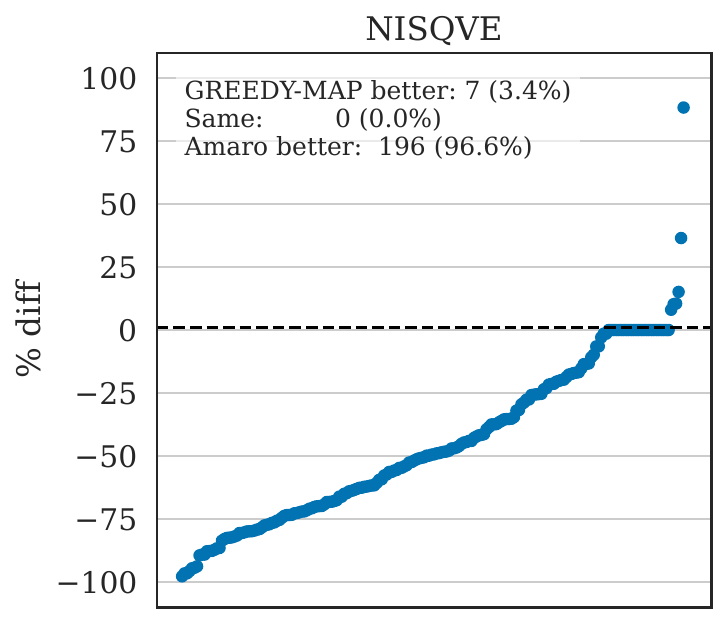}
\end{subfigure}%
\begin{subfigure}{0.23\linewidth}
  \includegraphics[width=\linewidth]{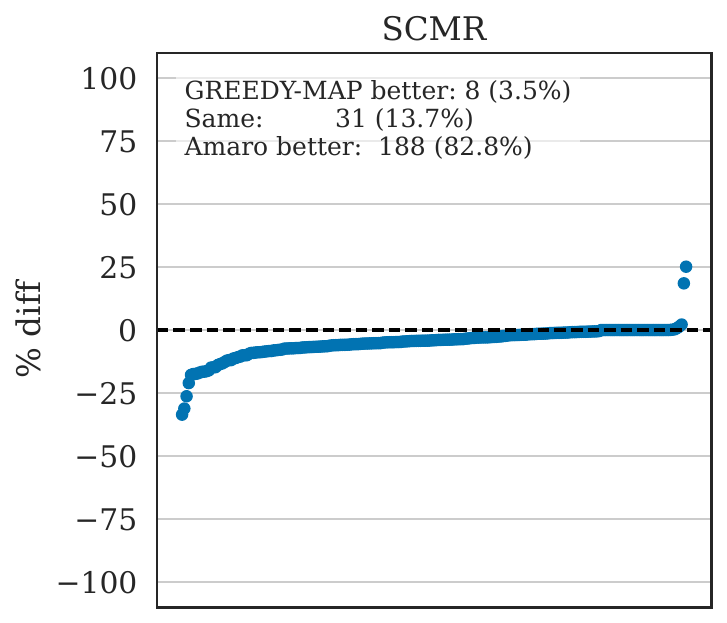}
\end{subfigure}
\begin{subfigure}{0.23\linewidth}
  \includegraphics[width=\linewidth]{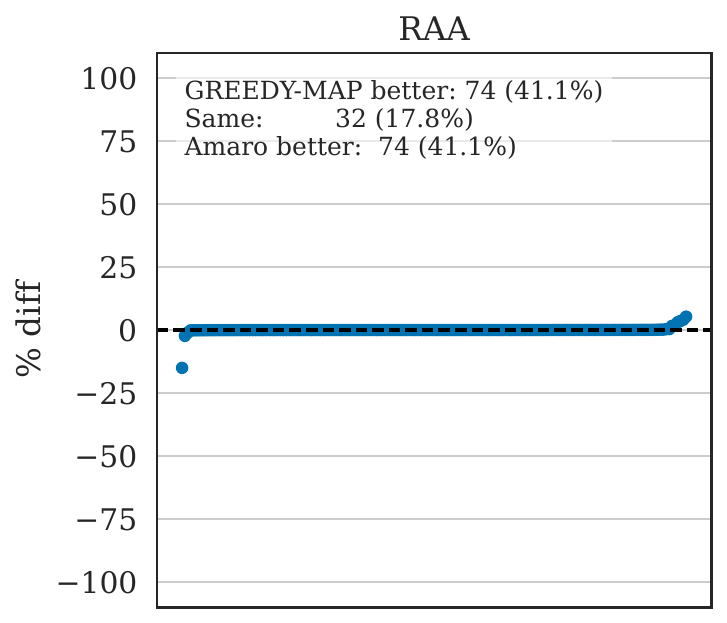}
\end{subfigure}
\begin{subfigure}{0.23\linewidth}
  \includegraphics[width=\linewidth]{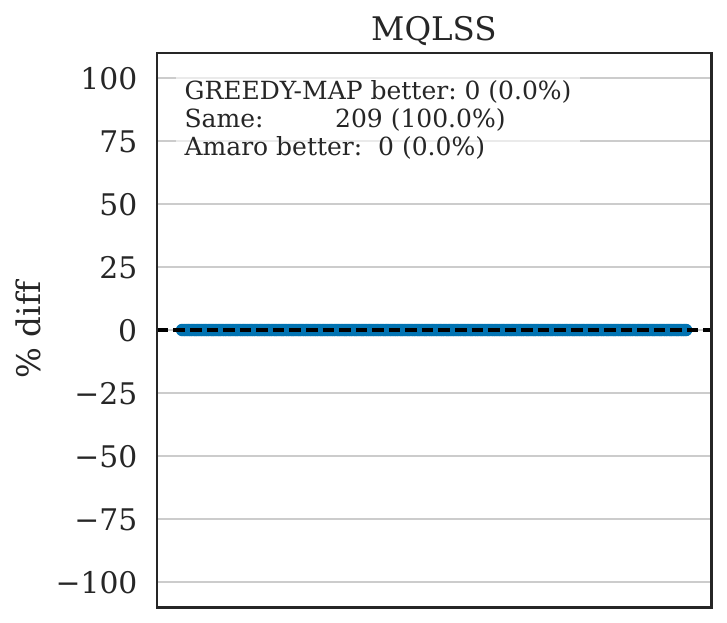}
\end{subfigure}
\begin{subfigure}{0.23\linewidth}
  \includegraphics[width=\linewidth]{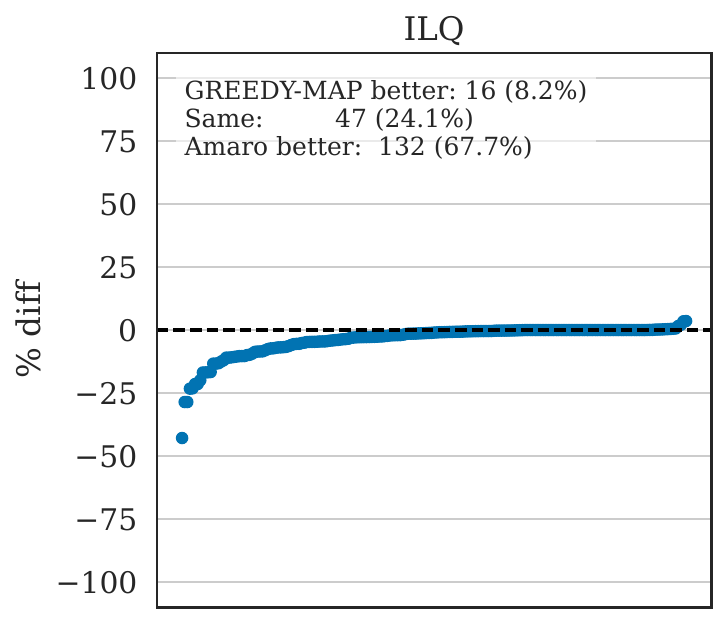}
\end{subfigure}
\begin{subfigure}{0.23\linewidth}
  \includegraphics[width=\linewidth]{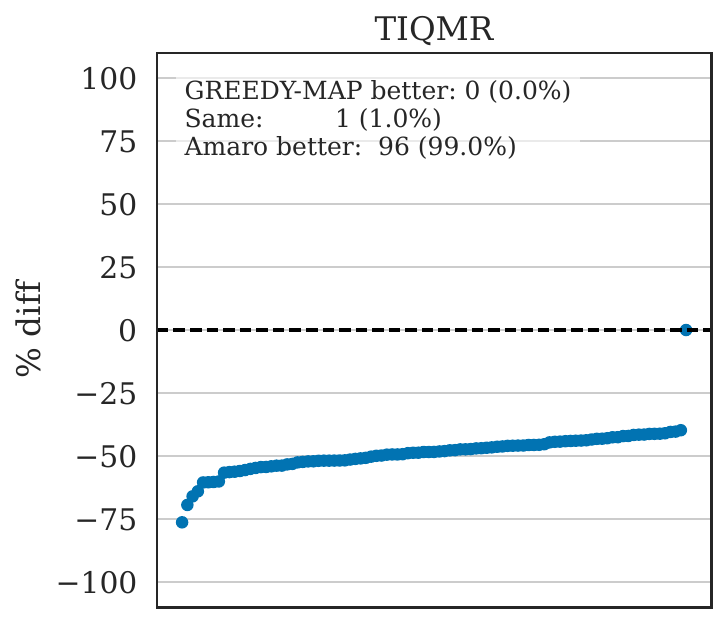}
\end{subfigure}
\caption{Initial map ablation}
\label{fig:map-ablation}
\end{figure}

\begin{figure}[h]
  \begin{subfigure}{0.23\linewidth}
  \includegraphics[width=\linewidth]{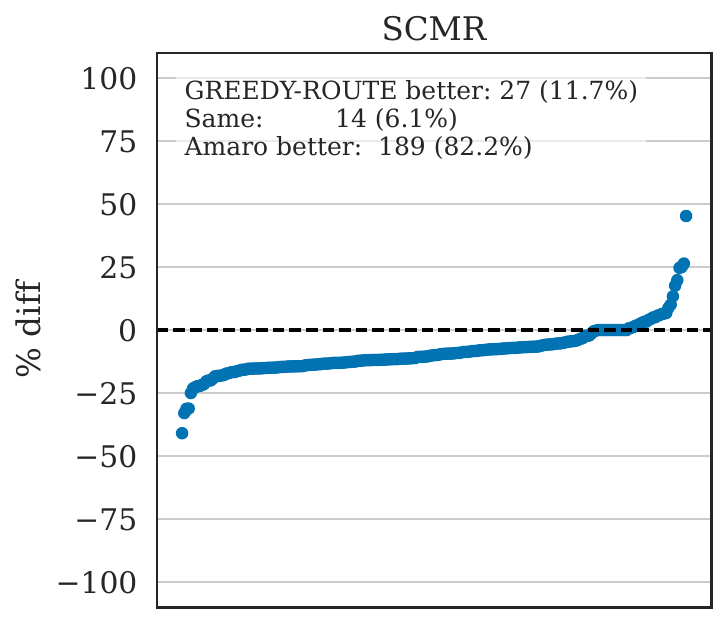}
\end{subfigure}
\begin{subfigure}{0.23\linewidth}
  \includegraphics[width=\linewidth]{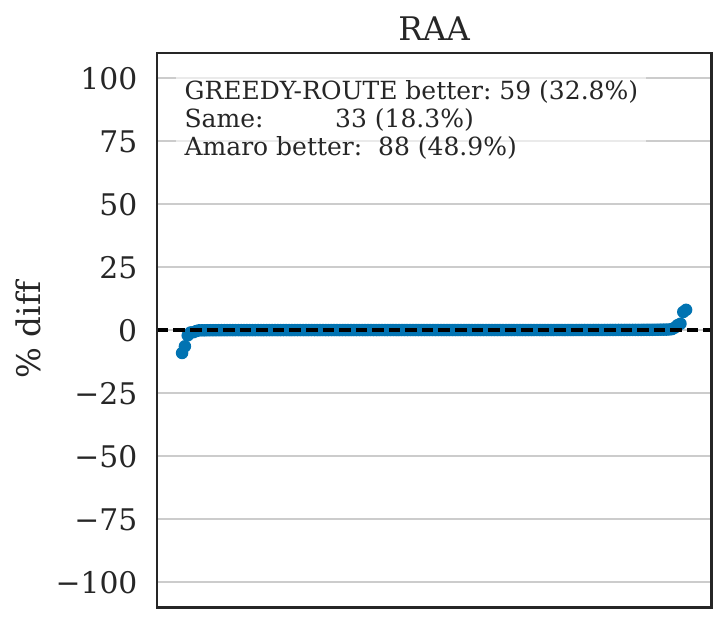}
\end{subfigure}
\begin{subfigure}{0.23\linewidth}
  \includegraphics[width=\linewidth]{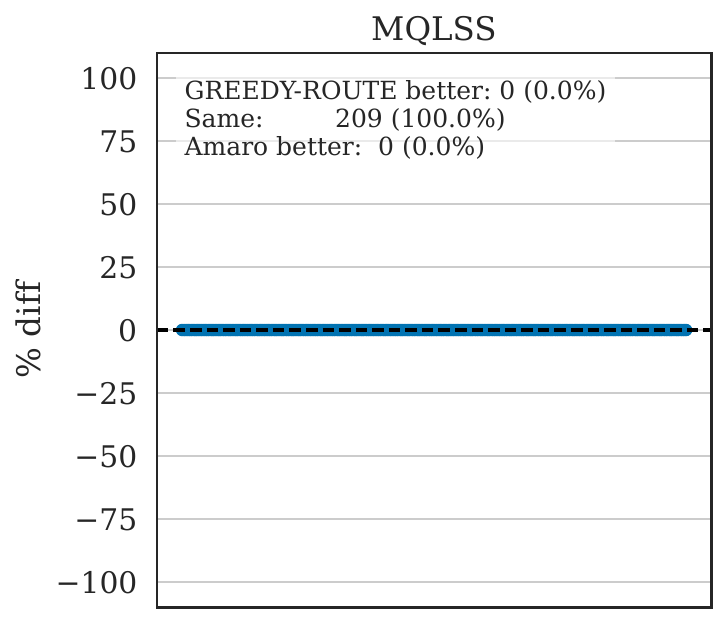}
\end{subfigure}
\begin{subfigure}{0.23\linewidth}
  \includegraphics[width=\linewidth]{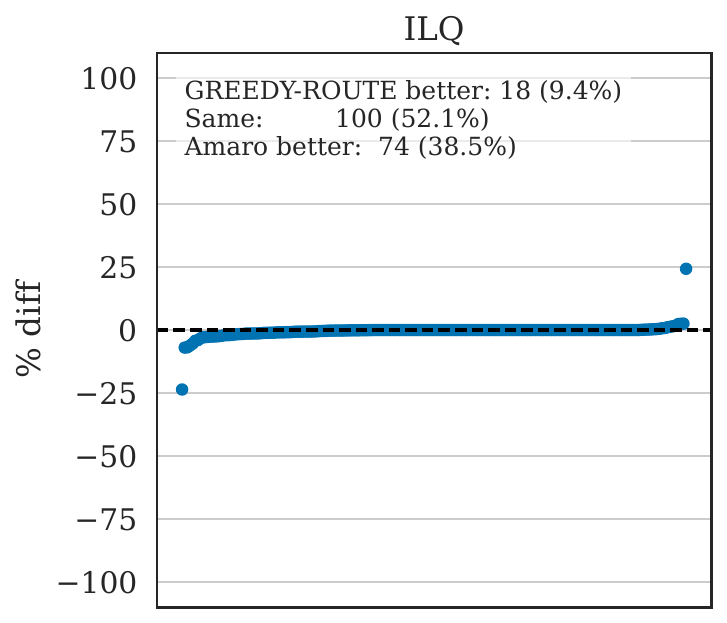}
\end{subfigure}
\caption{Maximal state search ablation}
\label{fig:route-ablation}
\end{figure}

\end{document}